\documentclass{aastex63}

\usepackage{multirow}
\let\tablenum\relax 
\usepackage{siunitx}
\usepackage{xspace}

\DeclareSIUnit{\Msol}{\ensuremath{M_\odot}}
\DeclareSIUnit{\mwe}{m.\,w.\,e.} 

\newcommand{\nue}{\ensuremath{\nu_e}\xspace}
\newcommand{\numu}{\ensuremath{\nu_\mu}\xspace}
\newcommand{\nutau}{\ensuremath{\nu_\tau}\xspace}
\newcommand{\nux}{\ensuremath{\nu_x}\xspace}
\newcommand{\nuebar}{\ensuremath{\bar{\nu}_e}\xspace}
\newcommand{\numubar}{\ensuremath{\bar{\nu}_\mu}\xspace}
\newcommand{\nutaubar}{\ensuremath{\bar{\nu}_\tau}\xspace}
\newcommand{\nuxbar}{\ensuremath{\bar{\nu}_x}\xspace}
\renewcommand{\d}{\mathrm{d}}

\newcommand\Tstrut{\rule{0pt}{2.6ex}}  
\newcommand\Bstrut{\rule[-1.2ex]{0pt}{0pt}}

\shorttitle{Supernova Model Discrimination with Hyper-Kamiokande}
\shortauthors{The Hyper-Kamiokande Collaboration}

\begin{document}

\title{Supernova Model Discrimination with Hyper-Kamiokande}

\newcommand{\BOSTON}{\affiliation{Boston University, Department of Physics, Boston, Massachusetts, USA}}
\newcommand{\UBC}{\affiliation{University of British Columbia, Department of Physics and Astronomy, Vancouver, British Columbia, Canada }}
\newcommand{\UCDAVIS}{\affiliation{University of California, Davis, Department of Physics, Davis, California, USA}}
\newcommand{\UCI}{\affiliation{University of California, Irvine, Department of Physics and Astronomy, Irvine, California, USA}}
\newcommand{\CSU}{\affiliation{California State University, Department of Physics, Carson, California, USA}}
\newcommand{\SACLAY}{\affiliation{IRFU, CEA, Universit\`e Paris-Saclay, Gif-sur-Yvette, France}}
\newcommand{\CHONNAM}{\affiliation{Chonnam National University, Department of Physics, Gwangju, Korea}}
\newcommand{\DONGSHIN}{\affiliation{Dongshin University, Laboratory for High Energy Physics, Naju, Korea}}
\newcommand{\LLR}{\affiliation{Ecole Polytechnique, IN2P3-CNRS, Laboratoire Leprince-Ringuet, Palaiseau, France}}
\newcommand{\LPNHE}{\affiliation{Laboratoire de Physique Nucleaire et de Hautes Energie, IN2P3/CNRS, Sorbonne Universit\`e, Paris, France}}
\newcommand{\EDINBURGH}{\affiliation{University of Edinburgh, School of Physics and Astronomy, Edinburgh, United Kingdom}}
\newcommand{\GENEVA}{\affiliation{University of Geneva, Section de Physique, DPNC, Geneva, Switzerland}}
\newcommand{\GIST}{\affiliation{GIST College, Gwangju Institute of Science and Technology, Gwangju, Korea}}
\newcommand{\HAWAII}{\affiliation{University of Hawaii, Department of Physics and Astronomy, Honolulu, Hawaii, USA}}
\newcommand{\IMPERIAL}{\affiliation{Imperial College London, Department of Physics, London, United Kingdom}}
\newcommand{\BARI}{\affiliation{INFN Sezione di Bari and Universit\`a e Politecnico di Bari,Bari Italy}}
\newcommand{\NAPOLI}{\affiliation{INFN Sezione di Napoli and Universit\`a Federico II di Napoli, Dipartimento di Fisica, Napoli, Italy}}
\newcommand{\PADOVA}{\affiliation{INFN Sezione di Padova and Universit\`a di Padova, Dipartimento di Fisica, Padova, Italy}}
\newcommand{\ROME}{\affiliation{INFN Sezione di Roma, Universit\`a Sapienza, Dipartimento di Fisica, Roma, Italy}}
\newcommand{\INR}{\affiliation{Institute for Nuclear Research of the Russian Academy of Sciences, Moscow, Russia}}
\newcommand{\ISU}{\affiliation{Iowa State University, Department of Physics and Astronomy, Ames, Iowa, USA}}
\newcommand{\KEK}{\affiliation{High Energy Accelerator Research Organization (KEK), Tsukuba, Japan}}
\newcommand{\JPARC}{\affiliation{J-PARC Center, Tokai, Japan}}
\newcommand{\SOKENDAI}{\affiliation{SOKENDAI (The Graduate University for Advanced Studies), Tokai, Japan}}
\newcommand{\KOBE}{\affiliation{Kobe University, Department of Physics, Kobe, Japan}}
\newcommand{\KYOTO}{\affiliation{Kyoto University, Department of Physics, Kyoto, Japan}}
\newcommand{\YITP}{\affiliation{Kyoto University, Yukawa Institute for Theoretical Physics, Kyoto, Japan}}
\newcommand{\LNF}{\affiliation{Laboratori Nazionali di Frascati, Frascati, Italy}}
\newcommand{\LANCASTER}{\affiliation{Lancaster University, Physics Department, Lancaster, United Kingdom}}
\newcommand{\LIVERPOOL}{\affiliation{University of Liverpool, Department of Physics, Liverpool, United Kingdom}}
\newcommand{\LANL}{\affiliation{Los Alamos National Laboratory, New Mexico, USA}}
\newcommand{\LSU}{\affiliation{Louisiana State University, Department of Physics and Astronomy, Baton Rouge, Louisiana, USA }}
\newcommand{\MADRID}{\affiliation{University Autonoma Madrid, Department of Theoretical Physics, Madrid, Spain}}
\newcommand{\MADRIDIFT}{\affiliation{UAM/CSIC, Instituto de F\'{\i}sica Te\'orica, Madrid, Spain}}
\newcommand{\MSU}{\affiliation{Michigan State University, Department of Physics and Astronomy,  East Lansing, Michigan, USA}}
\newcommand{\MIYAGI}{\affiliation{Miyagi University of Education, Department of Physics, Sendai, Japan}}
\newcommand{\NAGOYA}{\affiliation{Nagoya University, Graduate School of Science, Nagoya, Japan}}
\newcommand{\KMI}{\affiliation{Nagoya University, Kobayashi-Maskawa Institute for the Origin of Particles and the Universe, Nagoya, Japan}}
\newcommand{\STELAB}{\affiliation{Nagoya University, Institute for Space-Earth Environmental Research, Nagoya, Japan}}
\newcommand{\NCBJ}{\affiliation{National Centre for Nuclear Research, Warsaw, Poland}}
\newcommand{\OKAYAMA}{\affiliation{Okayama University, Department of Physics, Okayama, Japan}}
\newcommand{\OCU}{\affiliation{Osaka City University, Department of Physics, Osaka, Japan}}
\newcommand{\OXFORD}{\affiliation{Oxford University, Department of Physics, Oxford, United Kingdom}}
\newcommand{\PENN}{\affiliation{Pennsylvania State University, Department of Physics, University Park, Pennsylvania, USA}}
\newcommand{\PITTSBURGH}{\affiliation{University of Pittsburgh, Department of Physics and Astronomy, Pittsburgh, Pennsylvania, USA}}
\newcommand{\RAL}{\affiliation{STFC, Rutherford Appleton Laboratory, Harwell Oxford, and Daresbury Laboratory, Warrington, United Kingdom}}
\newcommand{\REGINA}{\affiliation{University of Regina, Department of Physics, Regina, Saskatchewan, Canada}}
\newcommand{\RHUL}{\affiliation{Royal Holloway University of London, Department of Physics, Egham, Surrey, United Kingdom}}
\newcommand{\RIO}{\affiliation{Pontif{\'\i}cia Universidade Cat{\'o}lica do Rio de Janeiro, Departamento de F\'{\i}sica, Rio de Janeiro, Brazil}}
\newcommand{\ROCHESTER}{\affiliation{University of Rochester, Department of Physics and Astronomy, Rochester, New York, USA}}
\newcommand{\RWTH}{\affiliation{RWTH Aachen University, III. Physikalisches Institut, Aachen, Germany}}
\newcommand{\QMUL}{\affiliation{Queen Mary University of London, School of Physics and Astronomy, London, United Kingdom}}
\newcommand{\SHEFFIELD}{\affiliation{University of Sheffield, Department of Physics and Astronomy, Sheffield, United Kingdom}}
\newcommand{\SNU}{\affiliation{Seoul National University, Department of Physics and Astronomy, Seoul, Korea}}
\newcommand{\SEOYEONG}{\affiliation{Seoyeong University, Department of Fire Safety, Gwangju, Korea }}
\newcommand{\STONYBROOK}{\affiliation{State University of New York at Stony Brook, Department of Physics and Astronomy, Stony Brook, New York, USA}}
\newcommand{\SKKU}{\affiliation{Sungkyunkwan University, Department of Physics, Suwon, Korea}}
\newcommand{\TOHOKU}{\affiliation{Tohoku University, Research Center for Neutrino Science, Sendai, Japan}}
\newcommand{\ERI}{\affiliation{University of Tokyo, Earthquake Research Institute, Tokyo, Japan}}
\newcommand{\KAMIOKA}{\affiliation{University of Tokyo, Institute for Cosmic Ray Research, Kamioka Observatory, Kamioka, Japan}}
\newcommand{\RCCN}{\affiliation{University of Tokyo, Institute for Cosmic Ray Research, Research Center for Cosmic Neutrinos, Kashiwa, Japan}}
\newcommand{\IPMU}{\affiliation{University of Tokyo, Kavli Institute for the Physics and Mathematics of the Universe (WPI), University of Tokyo Institutes for Advanced Study, Kashiwa, Japan}}
\newcommand{\TOKYO}{\affiliation{University of Tokyo, Department of Physics, Tokyo, Japan}}
\newcommand{\TODAI}{\affiliation{University of Tokyo, Tokyo, Japan}}
\newcommand{\TITECH}{\affiliation{Tokyo Institute of Technology, Department of Physics, Tokyo, Japan}}
\newcommand{\TRIUMF }{\affiliation{TRIUMF, Vancouver, British Columbia, Canada}}
\newcommand{\TORONTO}{\affiliation{University of Toronto, Department of Physics, Toronto, Ontario, Canada}}
\newcommand{\WARSAW}{\affiliation{University of Warsaw, Faculty of Physics, Warsaw, Poland}}
\newcommand{\WUT}{\affiliation{Warsaw University of Technology, Institute of Radioelectronics and Multimedia Technology, Warsaw, Poland}}
\newcommand{\WARWICK}{\affiliation{University of Warwick, Department of Physics, Coventry, United Kingdom}}
\newcommand{\WASHINGTON}{\affiliation{University of Washington, Department of Physics, Seattle, Washington, USA}}
\newcommand{\WINNIPEG}{\affiliation{University of Winnipeg, Department of Physics, Winnipeg, Manitoba, Canada}}
\newcommand{\VT}{\affiliation{Virginia Tech, Center for Neutrino Physics, Blacksburg, Virginia, USA}}
\newcommand{\WROCLAW}{\affiliation{Wroclaw University, Faculty of Physics and Astronomy, Wroclaw, Poland}}
\newcommand{\YEREVAN}{\affiliation{Institute for Theoretical Physics and Modeling, Yerevan, Armenia}}
\newcommand{\YORK}{\affiliation{York University, Department of Physics and Astronomy, Toronto, Ontario, Canada}}
\newcommand{\KYIV}{\affiliation{Kyiv National University, Department of Nuclear  Physics, Kyiv, Ukraine}}
\newcommand{\YOKOHAMA}{\affiliation{Yokohama National University, Faculty of Engineering, Yokohama, Japan}}
\newcommand{\TUS}{\affiliation{Tokyo University of Science, Department of Physics, Chiba, Japan}}
\newcommand{\STOCKHOLM}{\affiliation{Stockholm University, Oskar Klein Centre and Department of Physics, Stockholm, Sweden}}
\newcommand{\STOCKHOLMA}{\affiliation{Stockholm University, Oskar Klein Centre and Department of Astronomy, Stockholm, Sweden}}
\newcommand{\HCRI}{\affiliation{Harish-Chandra Research Institute, Allahabad, India}}
\newcommand{\IITG}{\affiliation{Indian Institute of Technology Guwahati, Guwahati, India}}
\newcommand{\TEZPUR}{\affiliation{Tezpur University, Department of Physics, Sonitpur, India}}
\newcommand{\GLASGOW}{\affiliation{University of Glasgow, School of Physics and Astronomy, Glasgow, United Kingdom}}
\newcommand{\VICTORIA}{\affiliation{University of Victoria, Department of Physics and Astronomy, Victoria, British Columbia, Canada}}
\newcommand{\CARLETON}{\affiliation{Carleton University, Department of Physics, Ottawa, Ontario, Canada}}
\newcommand{\BCIT}{\affiliation{British Columbia Institute of Technology, Physics Department, Burnaby, British Columbia,  Canada}}
\newcommand{\AGH}{\affiliation{AGH University of Science and Technology, Faculty of Computer Science, Electronics and Telecommunications, Krakow, Poland}}
\newcommand{\OAUJ}{\affiliation{Astronomical Observatory of the Jagiellonian University, Krakow, Poland}}
\newcommand{\KEIO}{\affiliation{Keio University, Department of Physics, Yokohama, Japan}}
\newcommand{\CANFRANC}{\affiliation{Laboratorio Subterr\'aneo de Canfranc, Canfranc-Estaci\'on, Spain}}
\newcommand{\CRACOW}{\affiliation{H. Niewodnicza\'nski Institute of Nuclear Physics PAN, Cracow, Poland}}
\newcommand{\SILESIA}{\affiliation{University of Silesia in Katowice, A. Che\l{}kowski Institute of Physics, Poland}}
\newcommand{\KCL}{\affiliation{King's College London, Department of Physics, Strand Building, Strand, London, United Kingdom}}
\newcommand{\UU}{\affiliation{Uppsala University, Department of Physics and Astronomy, Uppsala, Sweden}}
\newcommand{\GUADALAJARA}{\affiliation{Universidad de Guadalajara, CUCEI,  Departamento de Fisica, Guadalajara, Jal., Mexico}}
\newcommand{\GUADALAJARAI}{\affiliation{Universidad de Guadalajara, CUCEA, IT.Ph.D. program, Guadalajara, Jal., Mexico}}
\newcommand{\SINALOA}{\affiliation{Universidad Autonoma de Sinaloa, Culiacan, Mexico}}
\newcommand{\MOSCOW}{\affiliation{Moscow State University, Department of Theoretical Physics, Moscow, Russia}}
\newcommand{\SALERNOA}{\affiliation{Universit\`a degli Studi di Salerno and INFN Gruppo Collegato di Salerno, Fisciano, Italy}}
\newcommand{\SALERNOB}{\affiliation{INFN Gruppo Collegato di Salerno, Fisciano, Italy}}
\newcommand{\KTH}{\affiliation{KTH Royal Institute of Technology, Department of Physics, Stockholm, Sweden}}
\newcommand{\LNL}{\affiliation{INFN Laboratori Nazionali di Legnaro, Legnaro (PD), Italy}}
\newcommand{\VIIT}{\affiliation{Vishwakarma Institute of Information Technology, Pune, India}}  
\newcommand{\CAMPANIA}{\affiliation{Universit\`a della Campania ''L. Vanvitelli'' and INFN Sezione di Napoli, Napoli, Italy}} 
\newcommand{\INFNNA}{\affiliation{INFN Sezione di Napoli, Napoli, Italy}}
\newcommand{\KAIST}{\affiliation{Korea Institute of Science and Technology, Department of Physics, Daejeon, Korea}}
\newcommand{\TMU}{\affiliation{Tokyo Metropolitan University, Department of Physics, Tokyo, Japan}}
\newcommand{\UNIST}{\affiliation{Ulsan National Institute of Science and Technology, Department of Physics, Ulsan, Korea}}
\newcommand{\KNU}{\affiliation{Kyungpook National University, Department of Physics, Daegu, Korea}}
\newcommand{\CHARLES}{\affiliation{Charles University, IPNP, FMF, Prague, Czech}}
\newcommand{\IITJ}{\affiliation{Indian Institute of Technology Jodhpur, Department of Physics, Karwar, Rajasthan, India}}
\newcommand{\ETHZ}{\affiliation{ETH Zurich, Institute for Particle and Astroparticle Physics, Zurich, Switzerland}}
\newcommand{\SANGYO}{\affiliation{Kyoto Sangyo University, Department of Astrophysics and Atmospheric Sciences, Kyoto, Japan}}
\newcommand{\ITESM}{\affiliation{Tecnologico de Monterrey, Escuela de Ingenieria y Ciencias, Zapopan, Jalisco, Mexico}}
\newcommand{\LPI}{\affiliation{P.N.Lebedev Physical Institute of the Russian Academy of Sciences, Moscow, Russia}}
\newcommand{\NNSO}{\affiliation{University of Tokyo, Next-generation Neutrino Science Organization, Kamioka, Japan}}
\newcommand{\MOMA}{\affiliation{University of Oviedo, Applied Mathematical Modeling Group/Department of Physics, Oviedo, Spain}}
\newcommand{\DIPC}{\affiliation{Donostia International Physics Center and Ikerbasque Foundation, Basque Country, Spain}}
\newcommand{\VALENCIA}{\affiliation{Universitat Polit\`ecnica de Val\`encia, Instituto de Instrumentaci\`on para Imagen Molecular (i3M), Valencia, Spain}}
\newcommand{\ZARAGOZA}{\affiliation{University of Zaragoza, Centro de Astropart\'iculas y F\'isica de Altas Energ\'ias (CAPA), Zaragoza, Spain}} 
\newcommand{\SANTIAGO}{\affiliation{Universitat de Santiago de Compostela, Campus sur, Instituto Gallego de F\'isica de Altas Energ\'ias, Santiago de Compostela, Spain}}
\newcommand{\SNBOSE}{\affiliation{S.~N.~Bose National Centre for Basic Sciences, Salt Lake City, Kolkata, India}}

\AGH
\BOSTON
\UBC
\BCIT
\CAMPANIA
\CARLETON
\UCDAVIS
\UCI
\CSU
\CHARLES
\CHONNAM
\DONGSHIN
\DIPC
\LLR
\EDINBURGH
\ETHZ
\GENEVA
\GIST
\GLASGOW
\GUADALAJARA
\GUADALAJARAI
\CRACOW
\HAWAII
\IMPERIAL
\IITG
\IITJ
\SALERNOB
\LNL
\BARI
\NAPOLI
\INFNNA
\PADOVA
\ROME
\INR
\YEREVAN
\ISU
\SACLAY
\OAUJ
\JPARC
\KEIO
\KEK
\KCL
\KOBE
\KAIST
\KTH
\KYIV
\KYOTO
\SANGYO
\KNU
\LPNHE
\CANFRANC
\LANCASTER
\LIVERPOOL
\LANL
\LSU
\MADRID
\MADRIDIFT
\MSU
\MIYAGI
\ITESM
\MOSCOW
\NAGOYA
\STELAB
\KMI
\NCBJ
\STONYBROOK
\OKAYAMA
\OCU
\MOMA
\OXFORD
\LPI
\PENN
\PITTSBURGH
\VALENCIA
\RIO
\REGINA
\RWTH
\SALERNOA
\SANTIAGO
\SNU
\SEOYEONG
\SHEFFIELD
\SILESIA
\SINALOA
\SNBOSE
\SOKENDAI
\RAL
\STOCKHOLMA
\STOCKHOLM
\SKKU
\TEZPUR
\TOHOKU
\TODAI
\ERI
\KAMIOKA
\RCCN
\IPMU
\NNSO
\TOKYO
\TITECH
\TMU
\TUS
\TORONTO
\TRIUMF
\UNIST
\UU
\VICTORIA
\VIIT
\VT
\WARSAW
\WUT
\WARWICK
\WASHINGTON
\WINNIPEG
\WROCLAW
\YOKOHAMA
\YORK
\ZARAGOZA

\author{K.~Abe}\KAMIOKA\IPMU\NNSO
\author{P.~Adrich}\NCBJ
\author{H.~Aihara}\TOKYO\IPMU\NNSO
\author{R.~Akutsu}\TRIUMF
\author{I.~Alekseev}\LPI
\author{A.~Ali}\KYOTO
\author{F.~Ameli}\ROME
\author{I.~Anghel}\ISU
\author{L.H.V.~Anthony}\IMPERIAL
\author{M.~Antonova}\INR
% ...

\collaboration{9999}{Hyper-Kamiokande Collaboration}





\begin{abstract}

Core-collapse supernovae are among the most magnificent events in the observable universe.
They produce many of the chemical elements necessary for life to exist and their remnants---neutron stars and black holes---are interesting astrophysical objects in their own right.
However, despite millennia of observations and almost a century of astrophysical study, the explosion mechanism of core-collapse supernovae is not yet well understood.

Hyper-Kamiokande is a next-generation neutrino detector that will be able to observe the neutrino flux from the next galactic core-collapse supernova in unprecedented detail.
We focus on the first \SI{500}{ms} of the neutrino burst, corresponding to the accretion phase, and use a newly-developed, high-precision supernova event generator to simulate Hyper-Kamiokande’s response to five different supernova models.
We show that Hyper-Kamiokande will be able to distinguish between these models with high accuracy for a supernova at a distance of up to \SI{100}{kpc}.

Once the next galactic supernova happens, this ability will be a powerful tool for guiding simulations towards a precise reproduction of the explosion mechanism observed in nature.

\end{abstract}


\section{Introduction} \label{sec:intro}

A star with a mass of at least \SI{8}{\Msol} typically dies in a core-collapse supernova (ccSN).
In the process, large amounts of intermediate-mass chemical elements are created and ejected into interstellar space, influencing the star formation rate and stellar evolution in their galactic neighbourhood.
The compact remnant, meanwhile, is a neutron star or a black hole---important subjects of astrophysical research in their own right.
Understanding the ccSN explosion mechanism is therefore one of the central goals of astrophysics.

The electromagnetic emission from a ccSN begins minutes to hours after the initial explosion, when the outgoing shock wave breaks through the surface of the star\added{~\citep{Adams:2013ana}}.
It is therefore largely decoupled from the processes that occur during the explosion.
The observation of neutrinos from SN1987A was consistent with basic features predicted by the delayed neutrino-driven explosion mechanism developed by Wilson and Bethe in the 1980s~\citep{Wilson1982,Bethe1985}; specifically, the presence of an accretion phase in the first $\sim$\SI{500}{ms}~\citep{Loredo2002}.
However, with a total of two dozen events detected in the Kamiokande~\citep{Hirata1987,Hirata1988a}, IMB~\citep{Bionta1987} and Baksan~\citep{Alekseev1987} detectors, the available statistics were too low to determine details of the explosion mechanism.
\added{For a review of different analyses of these events, see~\citet{Vissani:2014doa}.}

In the decades since, progress in this area has largely relied on computer simulations.
While these simulations have made considerable progress and increasingly sophisticated three-dimensional models have become available in recent years (see e.\,g.~\citet{Hanke2013,OConnor2018,Burrows2020}), they are still limited by the available computing power and exhibit significant quantitative and in many cases even qualitative differences.

Once the next galactic ccSN happens, current and next-generation neutrino detectors will make a high-statistics observation of the neutrino burst \added{(see e.\,g.~\citet{Scholberg:2012id} for a review),} which will provide valuable input to simulations.
Previous work has demonstrated that this would make it possible to identify whether the signal exhibits certain features like \replaced{SASI}{the standing accretion shock instability (SASI)}~\citep{Lund2010,Tamborra2013} or \replaced{LESA}{lepton-number emission self-sustained asymmetry (LESA)}~\citep{Tamborra2014}\added{, or to characterize the stellar core e.\,g. by determining its compactness~\citep{Horiuchi2017} or the mass and radius of the resulting neutron star~\citep{Nakazato:2020ogl}}.
For recent reviews of expected features of the neutrino signal, see~\citet{Mirizzi:2015eza, Horiuchi2018}.
However, while model discrimination based on these features may allow us to exclude some classes of models, no general method for distinguishing between any two different ccSN models based on their neutrino signal has yet been presented.


In this paper, we present a log-likelihood method that makes optimal use of the full time and energy information available from many neutrino detectors to identify which supernova model best matches a set of observed events. 
Using a newly-developed, high-precision supernova event generator and a realistic detector simulation and event reconstruction, we investigate Hyper-Kamiokande’s response to five supernova models simulated by different groups around the world.
We show that this method requires just 100 (300) events within the first \SI{500}{ms} of the supernova burst---corresponding to a supernova distance of at least \SI{102}{kpc} (\SI{59}{kpc}) for normal mass ordering or \SI{97}{kpc} (\SI{56}{kpc}) for inverted mass ordering---to distinguish between different supernova models with (high) accuracy.

This paper is organized as follows.
We briefly introduce the Hyper-Kamiokande detector and summarise its sensitivity to supernova neutrinos in section~\ref{sec:hk}.
Section~\ref{sec:simulations} describes our simulations, giving an overview over the supernova models employed (section~\ref{sec:simulations-snmodels}), event generation (section~\ref{sec:simulations-sntools}), simulation and reconstruction (section~\ref{sec:simulations-simreco}) and data reduction (section~\ref{sec:simulations-cuts}).
In section~\ref{sec:results}, we present our likelihood function and determine how accurately it can distinguish between supernova models, before concluding in section~\ref{sec:summary}.

\section{Hyper-Kamiokande} \label{sec:hk}
Hyper-Kamiokande~\citep{HKDR2018} is a next-generation water Cherenkov detector that will be built near the town Kamioka in Japan’s Gifu Prefecture, approximately \SI{8}{km} south of the currently operating Super-Kamiokande detector~\citep{Fukuda2003}.
Located beneath the peak of Mount Nijugo, it will have an overburden of \SI{650}{m} of rock (\SI{1750}{\mwe}).
Construction has started in 2020, with data-taking scheduled to start in 2027.
Its physics goals include precision measurements of neutrino oscillation parameters (by measuring atmospheric neutrinos, accelerator neutrinos from the upgraded J-PARC beamline~\citep{Abe:2019fux} and solar neutrinos) as well as searches for proton decay and for astrophysical neutrinos from a wide range of sources.
In this section, we first give a brief overview over the detector design and then discuss Hyper-Kamiokande’s sensitivity to supernova neutrinos.

\subsection{Detector Design} \label{sec:hk-design}
\begin{figure}[htb]
	\centering
	\includegraphics[scale=0.4]{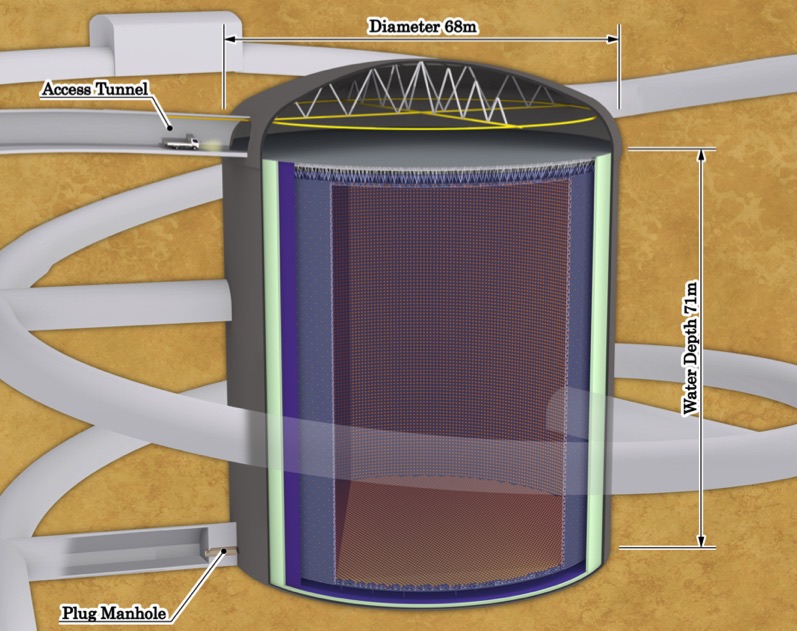}
	\caption{Drawing of the Hyper-Kamiokande detector.}
	\label{fig:hyperk}
\end{figure}
The basic design of Hyper-Kamiokande (see figure~\ref{fig:hyperk}) is similar to that of Super-Kamiokande.
It is a large, cylindrical detector with a height of \SI{71}{m} and diameter of \SI{68}{m}, filled with \SI{258}{kton} of ultra-pure water.\footnote{The simulations throughout this study used an earlier design with a height of \SI{60}{m} and diameter of \SI{74}{m}. The fiducial volume, and thus the results of this study, are not affected.}
It is optically separated into an outer detector with a width of \SI{2}{m} at the top and bottom or \SI{1}{m} at the sides, which acts as both shielding and active veto, and a \SI{217}{kton} cylindrical inner detector.
The structure dividing both detector regions has a diameter of \SI{60}{cm} and contains an array of photosensors as well as front-end electronics to collect and digitize signals from these photosensors.

The exact photosensor configuration of Hyper-Kamiokande has not yet been determined.
The baseline design uses a 40\,\% photocoverage with a new model of \SI{50}{cm} photomultiplier tube (PMT) that offers improved time and charge resolution compared to the PMT model used in Super-Kamiokande.
Alternative designs are currently being finalized.
Here, as a very conservative estimate, we assume a 20\,\% photocoverage with the new \SI{50}{cm} PMT model.

\subsection{Supernova Observations} \label{sec:hk-sn}
\begin{figure}[tb]
	\hspace{0.5pc}\begin{minipage}{22pc}
	\includegraphics[scale=0.43]{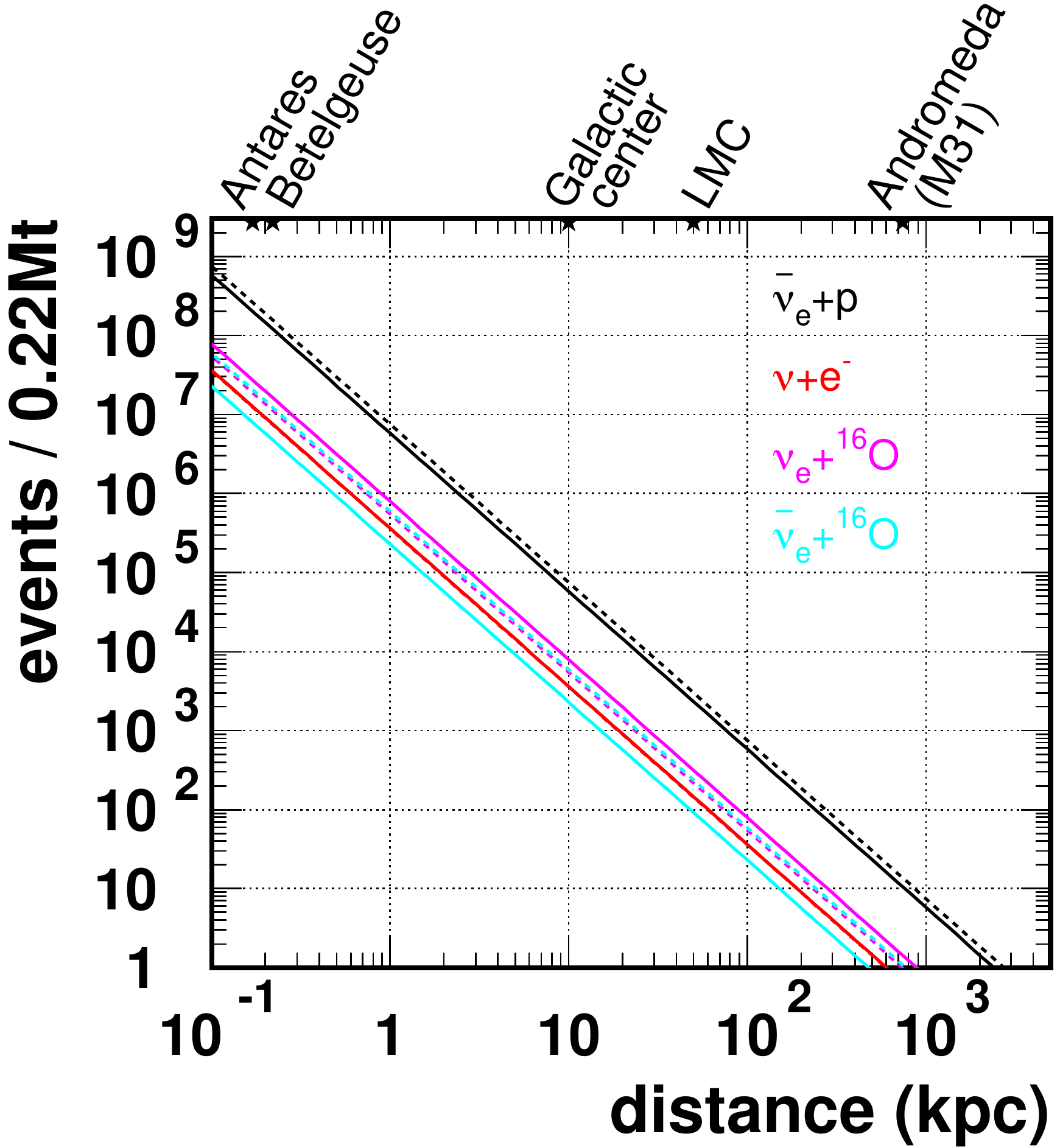}
	\end{minipage}
	\begin{minipage}{22pc}
	\vspace{1.05pc}
	\includegraphics[scale=0.39]{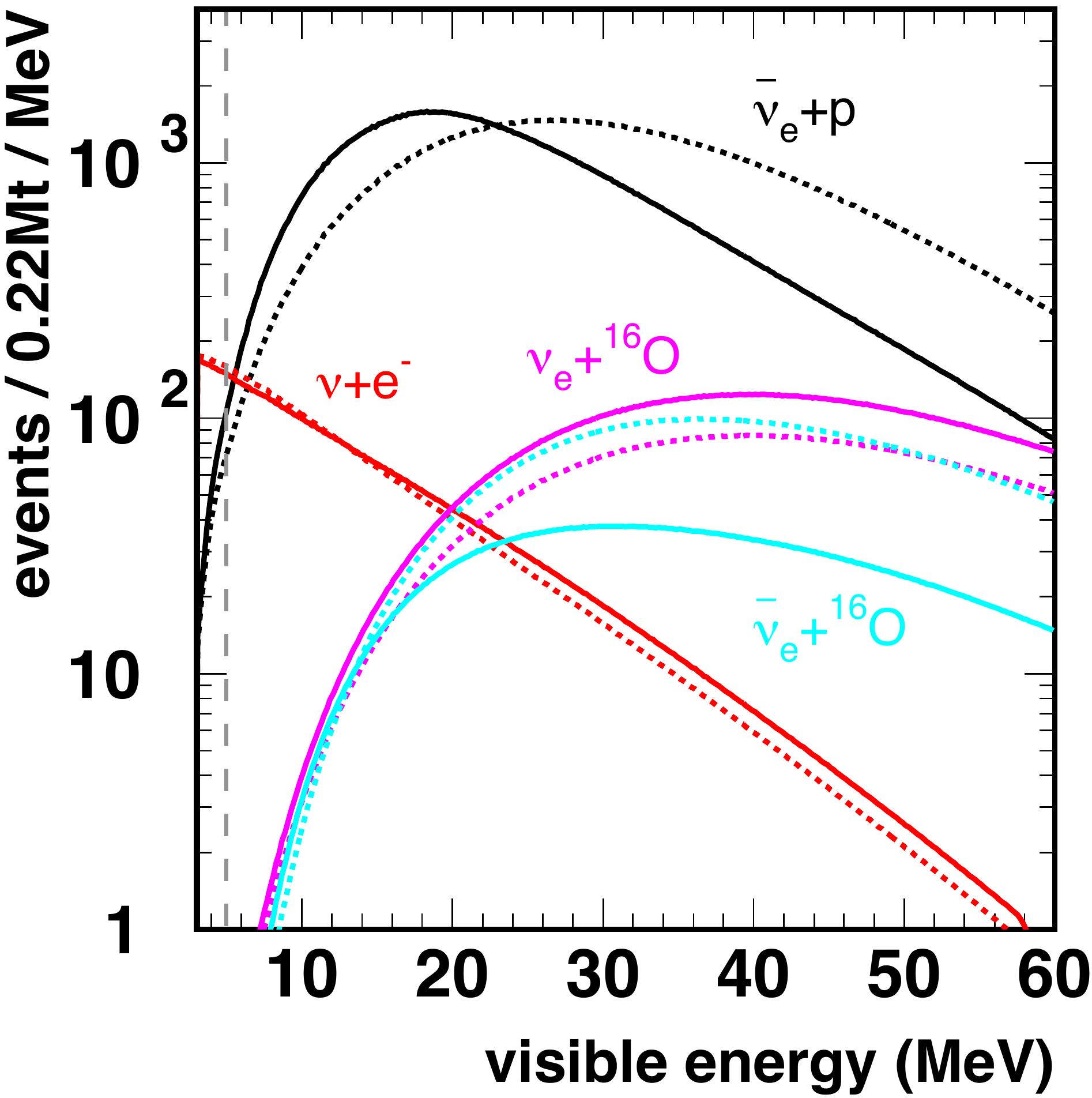}
	\end{minipage}
	\caption{Left: Expected number of events as a function of supernova distance. Right: True energy spectra of prompt events in the full inner detector for a supernova at \SI{10}{kpc}; for reference, the energy threshold used in this analysis (see section~\ref{sec:simulations-cuts}) is indicated by a dashed grey line. Both panels assume the supernova model by \citet{Totani1998}. Solid (dashed) lines correspond to normal (inverted) mass ordering, while different colours correspond to the interaction channels inverse beta decay (black), $\nu e$-scattering (red), \nue +$^{16}$O CC (purple) and \nuebar~+$^{16}$O CC (light blue).}
	\label{fig:evtsvdist_and_spectra}
\end{figure}

For a galactic supernova at a fiducial distance of \SI{10}{kpc}, Hyper-Kamiokande is expected to observe \numrange{54000}{90000} events in a burst with a total duration of a few tens of seconds.
For a nearby supernova (e.\,g. Betelgeuse at \SI{0.2}{kpc}), the peak event rate could reach \SI{e8}{Hz}. This rate was taken into account during the design of the DAQ system.
As shown in the left panel of figure~\ref{fig:evtsvdist_and_spectra}, the large volume also gives Hyper-Kamiokande an unprecedented ability to detect neutrinos from supernovae beyond the Milky Way:
For a supernova in the Large Magellanic Cloud at \SI{50}{kpc} distance, it would still detect about 3000 events, while for a supernova in the Andromeda galaxy (M31) at \SI{780}{kpc} distance, $\mathcal{O}(10)$ events are expected.

Hyper-Kamiokande can reconstruct the time and energy of each individual event, allowing it to reconstruct the neutrino spectrum.
The right panel of figure~\ref{fig:evtsvdist_and_spectra} shows energy spectra for the interaction channels considered in this paper.

The main interaction channel, inverse beta decay ($\nuebar + p \rightarrow n + e^+$), is responsible for about 90\,\% of events, making Hyper-Kamiokande most sensitive to \nuebar.
Elastic neutrino-electron scattering  ($\nu + e^- \rightarrow \nu + e^-$) is a subdominant interaction channel to which all neutrino flavours contribute.
The angular distribution of elastically scattered electrons is strongly peaked into a forward direction, which can be used to determine the direction of a supernova at the fiducial distance of \SI{10}{kpc} with an accuracy of about \SI{1}{\degree}~\citep{HKDR2018}.
Charged-current interactions of \nue and \nuebar on $^{16}$O nuclei are subdominant channels.
Due to their high energy threshold and the steep energy dependence of their cross sections, both channels are a very sensitive probe of the high-energy tail of the supernova neutrino flux, making up anywhere from $< 1\,\%$ to about 10\,\% of observed events.

\added{In this analysis, we focus on the prompt signal from the charged lepton in all interaction channels.
While Hyper-Kamiokande has some ability to detect, for example, the delayed neutron capture signal after inverse beta decay events, these events would be removed by the \SI{5}{MeV} energy cut introduced in section~\ref{sec:simulations-cuts}.
Similarly, we do not consider neutral-current interactions on $^{16}$O nuclei, a subdominant channel that mainly produces gamma rays with an energy of \SIrange{5.2}{6.3}{MeV}~\citep{Langanke:1995he}.
After Compton scattering on an electron or electron-positron pair production, the visible energy from these events would typically be below \SI{5}{MeV}.}

\section{Simulations} \label{sec:simulations}
\subsection{Supernova Models} \label{sec:simulations-snmodels}
While computer simulations of core-collapse supernovae have made significant advances in recent decades, they are still limited by the available computing power.
To overcome this problem, modelling groups employ a variety of different approximations and simplifying assumptions in their models, which lead to significant quantitative and in many cases even qualitative differences between different simulations.

Due to these uncertainties, and in order to demonstrate the broad applicability of the model discrimination method introduced in this work, we use a selection of five unrelated models here: a one-dimensional model that is primarily of historic interest (see section~\ref{sec:totani}), two one-dimensional models from recent parametric studies (see sections~\ref{sec:nakazato} and~\ref{sec:couch}) and two more complex multi-dimensional models (see sections~\ref{sec:tamborra} and~\ref{sec:vartanyan}).
These simulations were performed by different groups using a variety of progenitors and simulation codes.
They are intended to represent the much wider range of available models.

\begin{figure}[tbp]
	\hspace{1pc}\begin{minipage}{22pc}
	\includegraphics[scale=0.50]{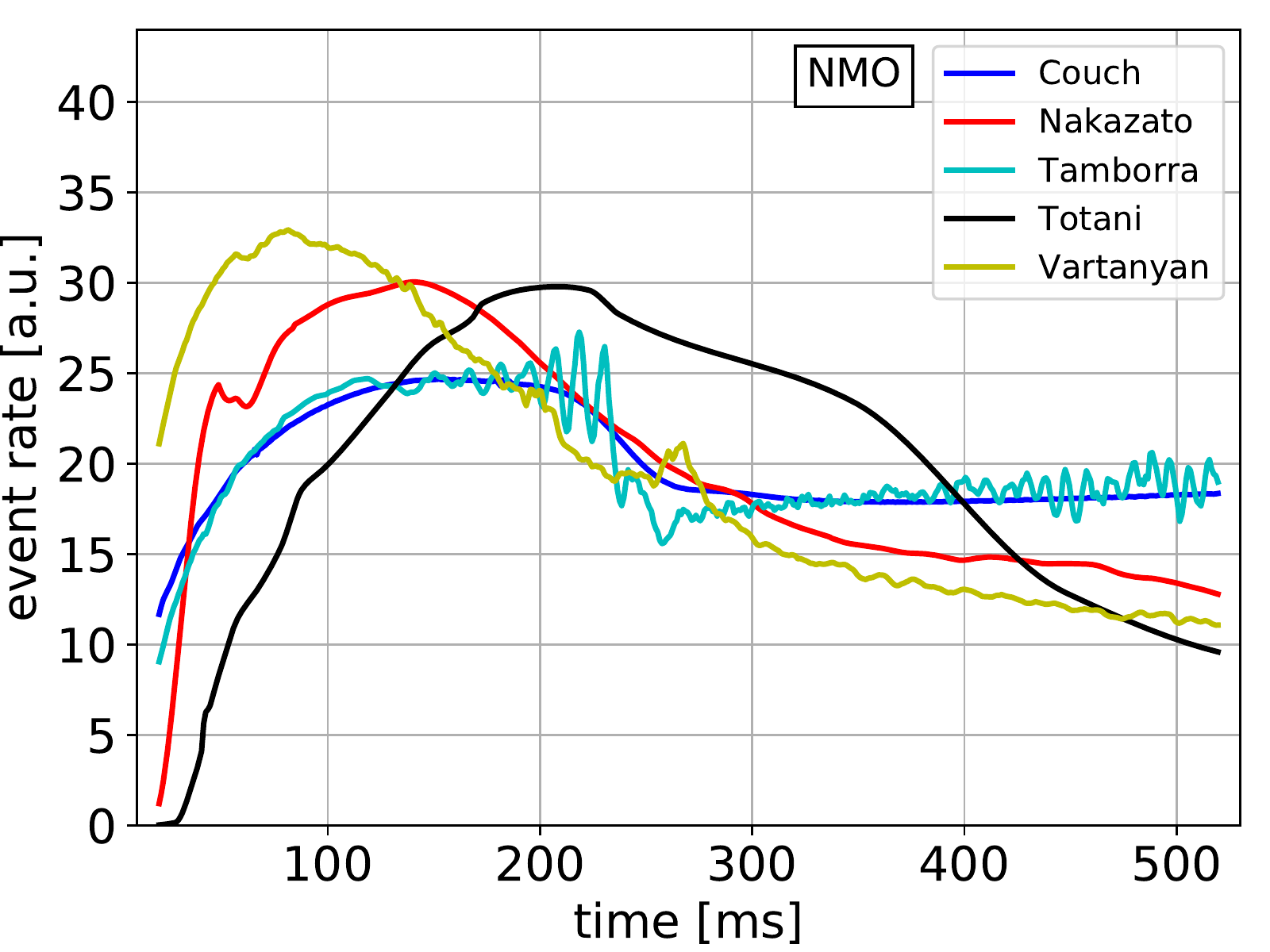}
	\end{minipage}
	\begin{minipage}{22pc}
	\includegraphics[scale=0.50]{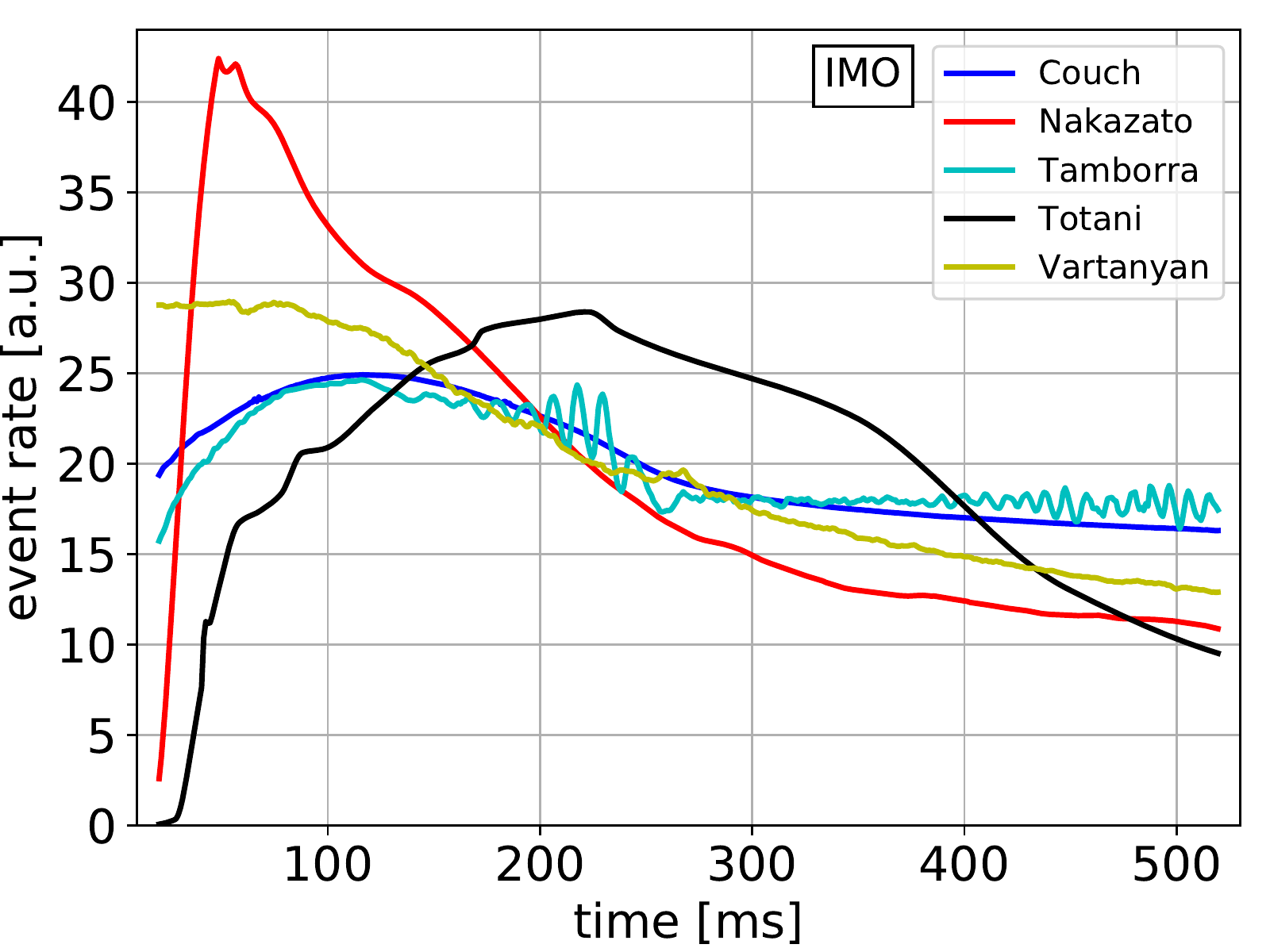}
	\end{minipage}

	\hspace{1pc}\begin{minipage}{22pc}
	\includegraphics[scale=0.50]{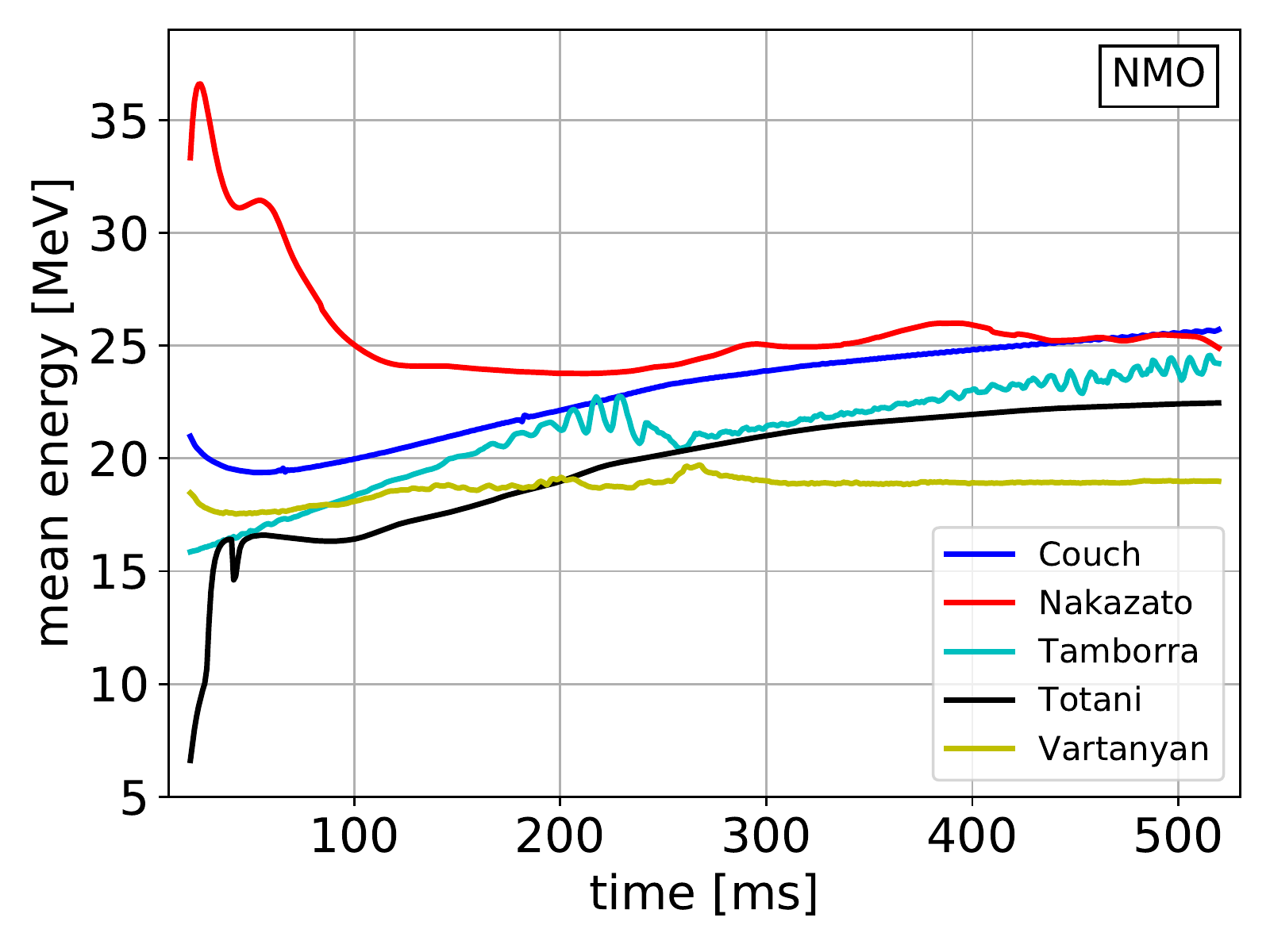}
	\end{minipage}
	\begin{minipage}{22pc}
	\includegraphics[scale=0.50]{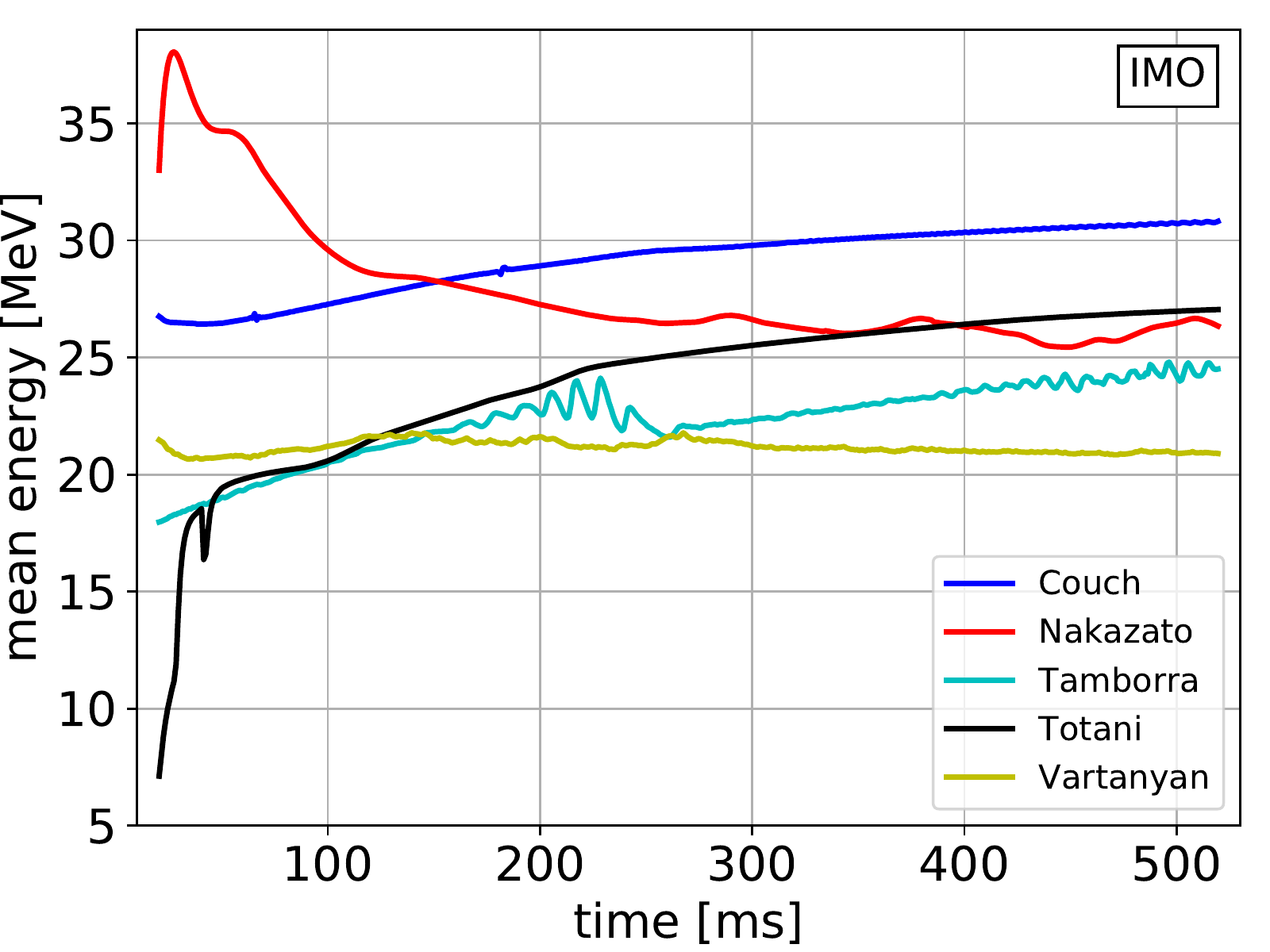}
	\end{minipage}
	\caption{Event rate (top) and mean energy (bottom) of observed events in Hyper-Kamiokande, as predicted by the five supernova models used in this paper for normal (left) or inverted (right) mass ordering. All plots show the time interval from \SIrange{20}{520}{ms} after core bounce. The event rate is normalized to produce the same total number of events for each model, reflecting the assumption made in this paper that the distance of the supernova is unknown.}
	\label{fig:simulations-models}
\end{figure}

Figure~\ref{fig:simulations-models} shows an overview over these models. In this section, we briefly describe these models.

\subsubsection{Totani}\label{sec:totani}
This one-dimensional model~\citep{Totani1998}, which is also referred to as the “Livermore model” or “Wilson model” in literature, is one of a small number of models that include the late-time evolution of the neutrino emission.
While it is now dated and has been surpassed by more accurate models, it is still used as a baseline model in many recent publications~\citep{Acciarri2015,Abe2016,HKDR2018}.

It uses a \SI{20}{\Msol} progenitor, which was modeled to reproduce the light curve of SN1987A, 
and a simulation code developed by Wilson and Mayle~\citep{Wilson1986,Mayle1987}.
Neutrino transport is modelled by the flux-limited diffusion approximation with 20 logarithmically spaced energy groups up to \SI{322.5}{MeV}.
The simulation is one-dimensional and was performed from the start of collapse to \SI{18}{s} after the core bounce.

\subsubsection{Nakazato}\label{sec:nakazato}
This family of models~\citep{Nakazato2013} contains progenitors with different initial masses and metallicities.
In this work, we focus on the \SI{20}{\Msol} progenitor with solar metallicity ($Z = 0.02$).
The one-dimensional simulation was performed from the start of collapse to \SI{20}{s} after the core bounce in two stages.
Here, we only use data from the first stage of the simulation, which contains the first \SI{520}{ms} post bounce.
It used the equation of state by~\citet{Shen1998}, 20 variably spaced energy groups up to \SI{300}{MeV} and a general relativistic neutrino-radiation-hydrodynamics ($\nu$RHD) code that solves the differential equations for hydrodynamics and neutrino transport simultaneously~\citep{Sumiyoshi2005}.


\subsubsection{Couch}\label{sec:couch}
This family of models~\citep{Couch2019,Warren2019} uses an approach for including effects of convection and turbulence in a one-dimensional simulation, which the authors call STIR (Supernova Turbulence In Reduced-dimensionality).
In this approach, the effective strength of convection depends on one parameter, $\alpha_\Lambda$, which can be tuned to reproduce results from a three-dimensional simulation of the same progenitor~\citep{OConnor2018}.
The model family contains 138 solar-metallicity progenitors with masses from \SIrange{9}{120}{\Msol}.
Here, we use results from the simulation of a \SI{20}{\Msol} progenitor~\citep{Sukhbold2014} with $\alpha_\Lambda = 0.8$.%
\footnote{This is an early version of the simulations, which differs slightly from the simulations described in the final version of~\citet{Couch2019}.}

The simulation was implemented in the FLASH simulation framework~\citep{Fryxell2000,Dubey2009} using a newly-implemented hydrodynamics solver 
with a modified effective potential to approximate effects of general relativity~\citep{Marek2006,OConnor2018b} and the SFHo equation of state~\citep{Steiner2013}.

Neutrino transport is simulated using a so-called “M1” transport scheme~\citep{OConnor2015,OConnor2018b} with 12 logarithmically spaced energy groups up to \SI{250}{MeV}.
Starting at \SI{5}{ms} post-bounce, effects of neutrino-electron scattering with energy transfer are turned off to reduce the computational resources required.
While this has little impact on the supernova dynamics, it does result in an increased mean energy for \nux~\citep{OConnor2018b}. 

\subsubsection{Tamborra}\label{sec:tamborra}
This model~\citep{Hanke2013,Tamborra2014} is a pioneering three-dimensional supernova simulation with energy-dependent neutrino transport.
We use results from the simulation of a \SI{27}{\Msol} progenitor~\citep{Woosley2002}.

The simulation was performed using the \textsc{Prometheus-Vertex} code consisting of the hydrodynamics solver \textsc{Prometheus}~\citep{Fryxell1991}, which implements the piecewise-parabolic method~\citep{Colella1984}, and the neutrino transport code \textsc{Vertex}~\citep{Rampp2002}, which uses the “ray-by-ray-plus” approach for velocity- and energy-dependent neutrino transport~\citep{Buras2006}.
The simulation uses a recent set of neutrino interaction rates~\citep{Muller2012}, the Lattimer and Swesty equation of state with compressibility $K = \SI{220}{MeV}$~\citep{Lattimer1991} and an effective potential to account for general relativistic corrections to Newtonian gravity~\citep{Marek2006}.
In multi-dimensional simulations, the neutrino signal inherently depends on the direction of the observer relative to the progenitor.
Here, we use the fluxes in the “violet” observer direction identified in~\citet{Tamborra2014}, which exhibits a large amplitude of the SASI oscillations in the luminosity and mean energy of neutrinos.

\subsubsection{Vartanyan}\label{sec:vartanyan}
This model is a recent two-dimensional simulation of a \SI{9}{\Msol} progenitor with solar metallicity~\citep{Sukhbold2016}.
It is similar to the simulations presented in~\citet{Radice2017, Seadrow2018} but used a different equation of state and grid resolution, which caused some physical and numerical differences.
As a result, while the luminosity and mean energy are qualitatively very similar to those described in~\citet{Seadrow2018}, exact values may differ by several percent.

This simulation was performed using the neutrino-radiation-hydrodynamics code \textsc{Fornax}~\citep{Skinner2019}, which combines a radiation hydrodynamics solver using a generalized variant of the piecewise-parabolic method~\citep{Colella1984} with neutrino transport using the “M1” scheme~\citep{Thorne1981,Shibata2011,Murchikova2017}.
It used 20 logarithmically spaced energy groups with energies up to \SI{300}{MeV} (\SI{100}{MeV}) for \nue (\nuebar and \nux), a detailed set of neutrino-matter interactions~\citep{Burrows2006}, the SFHo equation of state~\citep{Steiner2013} and an effective potential to account for general relativistic corrections to Newtonian gravity~\citep{Marek2006}.

\subsection{Event Generation} \label{sec:simulations-sntools}

We have developed a new supernova neutrino event generator called sntools~\replaced{\citep{sntools}}{\citep{Migenda2021}}.
It is open source\added{\footnote{\url{https://github.com/JostMigenda/sntools}}}, written in Python and makes heavy use of the NumPy~\citep{Walt2011} and SciPy~\citep{Virtanen2020} libraries.
In the following, we briefly discuss the neutrino interaction cross sections and treatment of flavour conversion implemented in sntools, before describing the generated data sets.

\subsubsection{Cross Sections}
sntools implements modern, high-precision cross-sections for the interaction channels described in section~\ref{sec:hk-sn}.
For inverse beta decay, it implements the full result from~\citet{Strumia2003}, including radiative corrections based on the approximation in~\citet{Kurylov2003}.
For neutrino-electron scattering, it implements the result from~\citet{Bahcall1995}, which includes one-loop QCD and electroweak corrections as well as QED radiative corrections.
For charged-current interactions of \nue and \nuebar on $^{16}$O, it implements a four-group fit~\citep{Nakazato2018} based on a recent shell model calculation~\citep{Suzuki2018}.

\subsubsection{Treatment of Neutrino Flavour Conversion}
As neutrinos produced inside the supernova traverse a smoothly varying density profile while exiting the star, they experience adiabatic flavour conversion via the MSW effect.
Afterwards, they propagate in a mass eigenstate until they interact.
The neutrino fluxes $\Phi_{\nu_i}$ observed in a detector are therefore linear combinations of the initial fluxes $\Phi_{\nu_i}^0$ predicted by the supernova simulation.
For normal mass ordering, this relation is given by~\citep{Dighe2000}
\begin{eqnarray}
\Phi_{\nue} &=& \sin^2 \theta_{13} \cdot \Phi_{\nue}^0 + \cos^2 \theta_{13} \cdot \Phi_{\nux}^0\\
\Phi_{\nuebar} &=& \cos^2 \theta_{12} \cos^2 \theta_{13} \cdot \Phi^0_{\nuebar} + (1 - \cos^2 \theta_{12} \cos^2 \theta_{13}) \cdot \Phi^0_{\nuxbar} \\
2 \Phi_{\nux} &=& \cos^2 \theta_{13} \cdot \Phi^0_{\nue} + (1 + \sin^2 \theta_{13}) \cdot \Phi^0_{\nux} \\
2 \Phi_{\nuxbar} &=& (1 - \cos^2 \theta_{12} \cos^2 \theta_{13}) \cdot \Phi^0_{\nuebar} + (1 + \cos^2 \theta_{12} \cos^2 \theta_{13}) \cdot \Phi^0_{\nuxbar},
\end{eqnarray}
while for inverted mass ordering, it is given by
\begin{eqnarray}
\Phi_{\nue} &=& \sin^2 \theta_{12} \cos^2 \theta_{13} \cdot \Phi_{\nue}^0 + (1 - \sin^2 \theta_{12} \cos^2 \theta_{13}) \cdot \Phi_{\nux}^0\\
\Phi_{\nuebar} &=& \sin^2 \theta_{13} \cdot \Phi_{\nuebar}^0 + \cos^2 \theta_{13} \cdot \Phi_{\nuxbar}^0\\
2 \Phi_{\nux} &=& (1 - \sin^2 \theta_{12} \cos^2 \theta_{13}) \cdot \Phi_{\nue}^0 + (1 + \sin^2 \theta_{12} \cos^2 \theta_{13}) \cdot \Phi_{\nux}^0\\
2 \Phi_{\nuxbar} &=& \cos^2 \theta_{13} \cdot \Phi_{\nuebar}^0 + (1 + \sin^2 \theta_{13}) \cdot \Phi^0_{\nuxbar}.
\end{eqnarray}

In both cases, the factor of 2 in the last two equations accounts for the fact that we use \nux (\nuxbar) to refer to either \numu or \nutau (either \numubar or \nutaubar), not to their sum.
These equations assume purely adiabatic transition (corresponding to $P_H = 0$ in~\citet{Dighe2000,Fogli2005}), which is appropriate during the early part of the neutrino emission we consider here~\citep{Fogli2005}.
For $\theta_{12}$ and $\theta_{13}$ we use values from the Particle Data Group~\citep{PDG2018}.
The effect of the uncertainty in both quantities on the generated data sets is much smaller than the random fluctuations between data sets generated from the same neutrino flux.

Neutrino self-interactions near the centre of the supernova could induce additional flavour conversion~\citep{Duan2006a,Duan2006}.
While these collective effects are the subject of intense theoretical study, no clear picture has yet emerged of how these effects will manifest in a given supernova~\citep{Chakraborty2016} and they are therefore not considered here.


\subsubsection{Data Sets}
Using sntools, we have generated data sets for the supernova models described in section~\ref{sec:simulations-snmodels}, for both normal and inverted mass ordering and for two different event counts per data set, as described below.
For every combination of these parameters, we have generated 1000 data sets in order to determine how accurately Hyper-Kamiokande is able to identify the true model despite random variations in each data set.
All events were distributed randomly within the inner detector of Hyper-Kamiokande.

\replaced{All data sets cover the time interval from \SIrange{20}{520}{ms} after core-bounce.
It contains the shock stagnation and accretion phase, which shows the largest differences between models.
The earlier (neutronization burst; see~\citet{Kachelries2005}) and later (proto-neutron star cooling; see~\citet{Suwa2019}) phases of neutrino emission are much better understood and exhibit only minor variations between models, making them less relevant for model discrimination.
}{
All data sets cover the time interval from \SIrange{20}{520}{ms} after core-bounce, which contains the shock stagnation and accretion phase.
The earlier neutronization burst is much better understood and exhibits only minor variations between models~\citep{Kachelries2005}, while the later diffusive proto-neutron star cooling is expected to be quasi-static and physically much simpler than the hydrodynamic behaviour of the shock wave~\citep{Suwa2019,Li:2020ujl}.
The accretion phase is thus expected to show the largest differences between models, making it most relevant for model discrimination.
}

Furthermore, due to the limited computing time available, many simulations---including the Couch, Vartanyan and Tamborra models used here---focus on the accretion phase and don’t include the full cooling phase.
Accordingly, by considering only this \SI{500}{ms} time interval we are able to include a wider range of models.

We have chosen the number of events per data set to be either 100 or 300.
The lower data set size was chosen in order to determine the lowest number of events needed to separate the models. 
As table~\ref{tab-ana-distance} shows, depending on the supernova model this correspond to a distance of at least \SI{102}{kpc} (\SI{97}{kpc}) for normal (inverted) mass ordering.
The larger size was chosen in order to demonstrate the increase in accuracy offered by a moderate increase in statistics.
300 events correspond to a supernova distance of at least \SI{59}{kpc} (\SI{56}{kpc}) for normal (inverted) mass ordering, and is thus representative of a supernova in the Large or Small Magellenic Cloud at a distance of \SI{50}{kpc}~\citep{Pietrzynski2013} or \SI{61}{kpc}~\citep{Hilditch2005}, respectively.
A much closer supernova, i.\,e. within the Milky Way, would offer a much higher event rate and thus even more accurate model discrimination.

\begin{table}[tbp]
\begin{center}
\begin{tabular}{lcccccc}
& \multicolumn{3}{c}{Normal Mass Ordering} & \multicolumn{3}{c}{Inverted Mass Ordering}\\
Model & $N_{\SI{10}{kpc}}$ & $d_{100}$ & $d_{300}$ & $N_{\SI{10}{kpc}}$ & $d_{100}$ & $d_{300}$\\
\hline
Totani & \num{20021} & \SI{141}{kpc} & \SI{82}{kpc} & \num{22717} & \SI{151}{kpc} & \SI{87}{kpc}\\
Nakazato & \num{17978} & \SI{134}{kpc} & \SI{77}{kpc} & \num{16005} & \SI{127}{kpc} & \SI{73}{kpc}\\
Couch & \num{27539} & \SI{166}{kpc} & \SI{96}{kpc} & \num{24983} & \SI{158}{kpc} & \SI{91}{kpc}\\
Vartanyan & \num{10372} & \SI{102}{kpc} & \SI{59}{kpc} & \num{9400} & \SI{97}{kpc} & \SI{56}{kpc}\\
Tamborra & \num{25025} & \SI{158}{kpc} & \SI{91}{kpc} & \num{20274} & \SI{142}{kpc} & \SI{82}{kpc}
\end{tabular}
\end{center}
\caption{Number of events expected during the time interval of \SIrange{20}{520}{ms} for a supernova at the fiducial distance of \SI{10}{kpc} ($N_{\SI{10}{kpc}}$) and the distances at which 100 or 300 events are expected in the inner detector of Hyper-Kamiokande ($d_{100}$ and $d_{300}$, respectively) for the five supernova models considered in this work and for both normal and inverted mass ordering.}
\label{tab-ana-distance}
\end{table}%
Throughout this analysis we assume that the distance to the supernova---and thus the normalization of the neutrino flux---is completely unknown; we only use the time and energy structure to distinguish between models.
If additional distance information is available---e.\,g. because an optical counterpart is identified---this could in principle be used to further improve the model discrimination accuracy. 

\subsection{Detector Simulation and Event Reconstruction} \label{sec:simulations-simreco}
To simulate events in Hyper-Kamiokande, we use WCSim~\citep{WCSim}, a package for simulating water Cherenkov detectors that is based on the physics simulation framework \textsc{Geant4}~\citep{Agostinelli2003} and the data analysis framework \textsc{root}~\citep{Brun1997}.
The vertex, direction and energy reconstruction follows the same approach developed by the Super-Kamiokande collaboration~\citep{Abe2011c} and is based on the BONSAI code~\citep{Smy2007}.
\added{Therefore, reconstruction performance is expected to be comparable to that achieved by Super-Kamiokande~\citep{Abe2016a}.}
As discussed in section~\ref{sec:hk-design}, the exact photosensor configuration of Hyper-Kamiokande has not yet been determined and we use a very conservative configuration with a 20\,\% photocoverage with \SI{50}{cm} PMTs.

\subsection{Data Reduction} \label{sec:simulations-cuts}

After reconstruction, we apply two cuts to all reconstructed events: an energy cut, which removes all events \replaced{with a reconstructed kinetic energy}{where the reconstructed kinetic energy of the detected $e^\pm$ is} less than \SI{5}{MeV}, and a fiducial volume cut, which removes all events whose reconstructed vertex is less than \SI{1.5}{m} away from the top, bottom or side walls of the inner detector.
The resulting fiducial mass is \SI{187}{kt}.

These cuts are intended to eliminate low-energy background from accidental coincidences of dark noise as well as radioactive decays in the detector.
Analogous cuts are also used for the solar neutrino analysis in Super-Kamiokande~\citep{Abe2016a}.
While several more advanced cuts used in that analysis, which rely on a comparison between MC simulations and observations, cannot currently be applied to the analysis presented here, the more stringent energy cut together with the much higher event rate (\SIrange{e2}{e3}{Hz} for the distant supernova bursts considered here, compared to about \SI{e-4}{Hz} for solar neutrinos in Super-Kamiokande) result in an effectively background-free data set.
Other backgrounds, including muon-induced spallation events or atmospheric neutrinos, occur at a much lower rate and are thus negligible during the single \SI{500}{ms} time interval considered here.

Once Hyper-Kamiokande is operating and the low-energy backgrounds are characterized in detail, it will likely be possible to develop more targeted cuts that allow us to include more low-energy events and extend the fiducial volume while remaining effectively background-free.


The fiducial volume cut described above removes about 13\,\% of all events in the inner detector.
The effect of the energy cut is generally small, though it depends on the energy spectrum of the initial neutrino flux and therefore on the supernova model and the mass ordering.
As an example, figure~\ref{fig:evtsvdist_and_spectra} shows the energy spectra in different interaction channels for the Totani model.
Due to the strong energy-dependence of the cross sections, three of the four interaction channels produce almost no events at \SI{5}{MeV} or below.
Only elastic $\nu e$-scattering---a subdominant channel which contributes about 5\,\% of all events---has a significant contribution at energies below \SI{5}{MeV}.
Overall, out of the initial 100 or 300 events per data set more than 80\,\% typically remain after applying these cuts.

\section{Results} \label{sec:results}
\subsection{Log-Likelihood Function} \label{sec:results-ll}

After the cuts described above, we apply an \added{unbinned} log-likelihood function to the reconstructed times and energies of the remaining events in each data set to determine how well that data set matches each of the supernova models.

\replaced{This log-likelihood function is similar to one that}{A similar function} was originally \replaced{derived}{used} for analysis of SN1987A taking into account only the main interaction channel, inverse beta decay~\citep{Loredo1989}.
However, the function used here includes all interaction channels.
It is derived in appendix~\ref{apx:likelihood} and given by
\begin{equation}
L = \ln \mathcal{L} = \sum_{i=1}^{N_\text{obs}} \ln \left( \sum_\alpha N_{i, \alpha} \right),
\end{equation}
where the index $i$ runs over the $N_\text{obs}$ events remaining in the data set and $N_{i, \alpha}$ is the number of events predicted by a given supernova model in the interaction channel $\alpha$ in an infinitesimally small bin around the reconstructed time and energy of event $i$.

By using infinitesimally small bins in time and energy, this likelihood function makes optimal use of all available information.
In contrast, using a binned chi-squared test to compare observation with models requires a sufficiently large number of events per bin to be accurate. 
Especially in the case of a distant supernova, where only hundreds or thousands of events may be observed in Hyper-Kamiokande, two-dimensional binning in time and energy would only be possible in very coarse bins, which would lose a lot of the available information.

The absolute numerical values of this likelihood function depend on the bin size chosen and are therefore not physically meaningful.
However, when calculating likelihood ratios for different models (i.\,e. differences in the log-likelihood, $\Delta L = L_A - L_B$), this dependence cancels out and the ratio describes whether model A or B is more likely to produce a given data set.
We will therefore exclusively use likelihood ratios to compare different models.

\subsection{N=100 Events per Data Set} \label{sec:results-100}

\begin{figure}[tbp]
	\centering
	\includegraphics[scale=0.75]{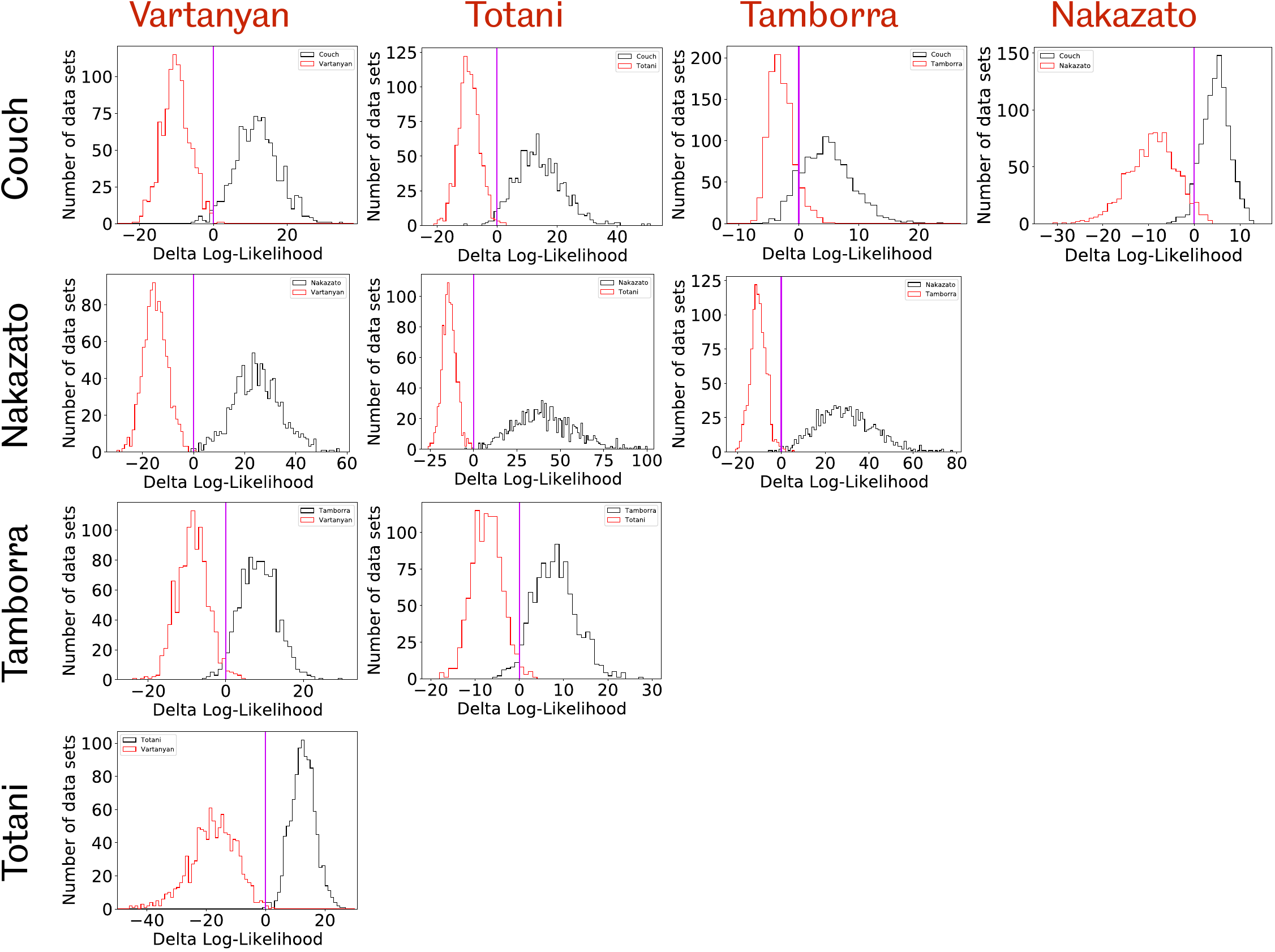}
	\caption{Histograms showing the distribution of $\Delta L = L_\text{black} - L_\text{red}$ for all pairs of supernova models considered here, for 100 events per data set, normal mass ordering and 20\,\% photocoverage. The purple vertical line in each panel indicates $\Delta L = 0$.}
	\label{fig-ana-100normal}
\end{figure}

Figure~\ref{fig-ana-100normal} shows pairwise comparisons of the five different models described in section~\ref{sec:simulations-snmodels} for normal mass ordering.
For example, the top right panel shows a comparison of the Couch model (black histogram) and Nakazato model (red histogram).
For most data sets generated from the Couch (Nakazato) model, $\Delta L = L_\text{Couch} - L_\text{Nakazato}$ is positive (negative), indicating that this method is generally able to identify the true model.
However, the overlap of both histograms indicates that misidentification sometimes occurs because of random fluctuations in the data sets.
Other model pairs in the figure show a similarly clear separation with only minor overlap around $\Delta L = 0$.
The largest overlap is seen between the Couch and Tamborra models, indicating that these models are most similar and hardest to distinguish.

This can be seen more clearly in 
the top half of table~\ref{tab-ana-100}, which compares all five supernova models simultaneously by determining which of them produces the highest likelihood for a given data set.
For each model, the respective row indicates what fraction of the 1000 generated data sets were identified as which model.
For example, 85.3\,\% of Tamborra data sets were identified correctly, while 8.4\,\% were misidentified as corresponding to the Couch model.
For the Couch model, almost 80\,\% of data sets were identified correctly, with most of the remaining data sets being misidentified as the Tamborra model.
Finally, the three other models are identified correctly in over 95\,\% of all cases.

The bottom half of table~\ref{tab-ana-100} shows results for the inverted mass ordering.
In this scenario, the largest overlap is observed between the Tamborra and Vartanyan models, with an 85--90\,\% chance of identifying those data sets correctly and a chance of just over 10\,\% of confusing these models for one another.
As for the normal mass ordering, the other three models are identified correctly in over 95\,\% of all cases.
Histograms showing one-to-one comparisons for each pair of models can be found in appendix~\ref{apx:model-comparisons}.


\begin{table}[tbp]
\begin{center}
\begin{tabular}{cl|rrrrr}
\multicolumn{2}{c}{~} & \multicolumn{5}{c}{\textbf{Reconstructed Model}}\\
\multicolumn{2}{r|}{\textbf{Normal}} & Couch & Nakazato & Tamborra & Totani & Vartanyan\\
\cline{2-7}
\multirow{5}{*}{\rotatebox{90}{\Tstrut \textbf{True Model\,}\Bstrut}} &
Couch & \textbf{79.5} & 5.7 & 12.2 & 1.2 & 1.4\\
& Nakazato & 3.3 & \textbf{96.1} & 0.3 & 0.1 & 0.2\\
& Tamborra & 8.4 & 0.0 & \textbf{85.3} & 3.3 & 3.0\\
& Totani & 0.4 & 0.0 & 1.6 & \textbf{97.9} & 0.1\\
& Vartanyan & 0.0 & 0.1 & 1.7 & 0.3 & \textbf{97.9}
\end{tabular}
\end{center}

\begin{center}
\begin{tabular}{cl|rrrrr}
\multicolumn{2}{c}{~} & \multicolumn{5}{c}{\textbf{Reconstructed Model}}\\
\multicolumn{2}{r|}{\textbf{Inverted}} & Couch & Nakazato & Tamborra & Totani & Vartanyan\\
\cline{2-7}
\multirow{5}{*}{\rotatebox{90}{\Tstrut \textbf{True Model\,}\Bstrut}} &
Couch & \textbf{96.0} & 3.5 & 0.4 & 0.1 & 0.0\\
& Nakazato & 0.8 & \textbf{99.2} & 0.0 & 0.0 & 0.0\\
& Tamborra & 0.0 & 0.1 & \textbf{85.8} & 2.1 & 12.0\\
& Totani & 0.3 & 0.0 & 2.0 & \textbf{97.7} & 0.0\\
& Vartanyan & 0.0 & 0.2 & 10.5 & 0.1 & \textbf{89.2}
\end{tabular}
\end{center}
\caption{Accuracy with which the true model can be identified, for 100 events per data set. Each line shows what fraction (in \%) of the 1000 data sets generated for a given model (left column) were identified as each of the five models. Correctly identified models are \textbf{highlighted}. Top: Normal mass ordering. Bottom: Inverted mass ordering.}
\label{tab-ana-100}
\end{table}%

\subsection{N=300 Events per Data Set} \label{sec:results-300}

When considering larger data sets, the effect of random fluctuations between individual data sets will decrease.
As a result, the accuracy of model identification is expected to increase significantly.

Table~\ref{tab-ana-300} shows results of the model identification for 300 events per data set, which are consistent with this expectation.
The top half shows results for normal mass ordering. The Couch and Tamborra models---which are most likely to be confused for each other in normal mass ordering---are now identified correctly with about 98\,\% accuracy and the probability of misidentifying one for the other is just 1.6\,\%.
The bottom half shows results for inverted mass ordering. The Tamborra and Vartanyan models---which are most likely to be confused for each other in inverted mass ordering---are now identified correctly with over 97\,\% accuracy.

For both normal and inverted mass ordering, the other three models are identified correctly with at least 99.9\,\% accuracy.
Histograms showing one-to-one comparisons for each pair of models can be found in appendix~\ref{apx:model-comparisons}.

\begin{table}[tbp]
\begin{center}
\begin{tabular}{cl|rrrrr}
\multicolumn{2}{c}{~} & \multicolumn{5}{c}{\textbf{Reconstructed Model}}\\
\multicolumn{2}{r|}{\textbf{Normal}} & Couch & Nakazato & Tamborra & Totani & Vartanyan\\
\cline{2-7}
\multirow{5}{*}{\rotatebox{90}{\Tstrut \textbf{True Model\,}\Bstrut}} &
Couch & \textbf{98.2} & 0.2 & 1.6 & 0.0 & 0.0\\
& Nakazato & 0.1 & \textbf{99.9} & 0.0 & 0.0 & 0.0\\
& Tamborra & 1.6 & 0.0 & \textbf{98.0} & 0.2 & 0.2\\
& Totani & 0.0 & 0.0 & 0.0 & \textbf{100.0} & 0.0\\
& Vartanyan & 0.0 & 0.0 & 0.0 & 0.0 & \textbf{100.0}
\end{tabular}
\end{center}

\begin{center}
\begin{tabular}{cl|rrrrr}
\multicolumn{2}{c}{~} & \multicolumn{5}{c}{\textbf{Reconstructed Model}}\\
\multicolumn{2}{r|}{\textbf{Inverted}} & Couch & Nakazato & Tamborra & Totani & Vartanyan\\
\cline{2-7}
\multirow{5}{*}{\rotatebox{90}{\Tstrut \textbf{True Model\,}\Bstrut}} &
Couch & \textbf{99.9} & 0.1 & 0.0 & 0.0 & 0.0\\
& Nakazato & 0.0 & \textbf{100.0} & 0.0 & 0.0 & 0.0\\
& Tamborra & 0.0 & 0.0 & \textbf{97.4} & 0.1 & 2.5\\
& Totani & 0.0 & 0.0 & 0.0 & \textbf{100.0} & 0.0\\
& Vartanyan & 0.0 & 0.0 & 0.8 & 0.0 & \textbf{99.2}
\end{tabular}
\end{center}
\caption{Same as table~\ref{tab-ana-100} but for 300 events per data set. Top: Normal mass ordering. Bottom: Inverted mass ordering.}
\label{tab-ana-300}
\end{table}%

\subsection{Observation of an Actual Supernova Neutrino Burst} \label{sec:bayes}

Above, we have answered the following question: Assuming that model X describes the true neutrino fluxes from a supernova, how likely are we to correctly identify X when comparing it with a range of other models?
This lets us identify which models are more or less similar to each other and assess Hyper-Kamiokande’s model discrimination capabilities.
However, it does not reflect the scenario we will face in the future when we observe a single supernova neutrino burst and do not know the true model.

Thus, another question of interest is: Assuming that we observe a supernova neutrino burst that is best described by model X, how confident are we that we can exclude some alternative model Y?
To answer this, we need to consider the interpretation of the likelihood ratio.

In a Bayesian interpretation~\citep{Loredo2002}, the ratio of likelihoods for two models A and B is equal to the Bayes factor $B_\text{AB}$ and equivalently, the difference in log-likelihoods is $\Delta L = \ln B_\text{AB}$.
If there is no \emph{a priori} reason to prefer one model over the other, this can be used to exclude disfavoured models beyond a certain threshold.

\begin{table}[btp]
\begin{center}
\begin{tabular}{lll}
$\ln B_\text{AB}$ & $B_\text{AB}$ & Evidence for model A over model B\\
\hline
0 to 1 & 1 to 3 & Negligible\\
1 to 3 & 3 to 20 & Positive\\
3 to 5 & 20 to 150 & Strong\\
$>5$ & $>150$ & Very strong
\end{tabular}
\end{center}
\caption{Interpretation of Bayes factor when comparing two models A and B. Adapted from~\citet{Kass1995}.}
\label{tab-ana-bayes}
\end{table}%

A suggested interpretation of Bayes factors is listed in table~\ref{tab-ana-bayes}.%
\footnote{Note that we show $B_\text{ij}$ here, whereas the original paper lists $2 B_\text{ij}$ due to its similarity with the more familiar $\Delta \chi^2$ values.~\citep{Kass1995,Ianni2009}}
Looking at the pairwise model comparison in figure~\ref{fig-ana-100normal}, 
we see that this interpretation matches our intuition:
The range from $\Delta L = -5$ to 5 contains almost the complete overlap between both histograms, where misidentification of data sets may occur, indicating that requiring $\Delta L \geq -5$ is unlikely to wrongly exclude the true model.
At the same time, most data sets based on the wrong model are correctly excluded by this criterion.
Once we observe an actual supernova neutrino burst, this criterion will therefore allow us to narrow down the list of supernova models that are compatible with the observed signal.

Since only likelihood ratios are physically meaningful as discussed in section~\ref{sec:results-ll}, this will determine which model fits the observed events better than all other models.
To determine whether the preferred model is actually compatible with the data, a separate goodness-of-fit test is required.

\section{Summary and Discussion} \label{sec:summary}

In this study, we introduced a likelihood function to determine how well neutrino fluxes predicted by a supernova model match an observed set of events.
It makes optimal use of the timing and energy information of every reconstructed event and includes four interaction channels relevant for water Cherenkov detectors.
This method is highly versatile and can in principle be used to determine any factor that affects the neutrino flux from a supernova.

As a proof of principle, we selected five different supernova models and generated data sets of 100 (300) events in Hyper-Kamiokande, corresponding to a supernova distance of at least \SI{102}{kpc} (\SI{59}{kpc}) for normal mass ordering or \SI{97}{kpc} (\SI{56}{kpc}) for inverted mass ordering.
We then simulated and reconstructed these events with the experiment’s official software toolchain.
For both normal and inverted mass ordering, using this method lets us identify the correct supernova model with high accuracy.

When the next supernova happens in the Milky Way or one of the nearby dwarf galaxies, Hyper-Kamiokande will thus be able to reliably identify a small number of supernova models that best match the observed neutrino burst.
This will be a powerful tool for guiding models towards a precise reproduction of the explosion mechanism observed in nature.

Throughout this study, we have assumed a very conservative detector configuration with a 20\,\% photocoverage.
An increased photocoverage would improve the detector performance, particularly at low energies, which may allow us to introduce more targeted cuts and include more signal events while remaining effectively background-free.
With improved low-energy performance, it would also be possible to tag neutrino interactions producing neutrons by detecting \SI{2.2}{MeV} gamma-rays from neutron capture on hydrogen.
This would let us distinguish between different interaction channels on an event-by-event basis and determine the fluxes of neutrinos and antineutrinos separately, which could further improve our model discrimination accuracy~\citep{Nikrant2018}.

We have also assumed that the distance of the supernova is unknown, which leaves the normalization of the supernova fluxes open.
If the distance of the supernova can be determined to sufficient accuracy---e.\,g. if an optical counterpart is visible---this would fix that normalization factor and further help distinguish between models that predict a different number of events at a fixed distance.
Other neutrino experiments---particularly those employing different and complementary detection techniques---could apply an analogous likelihood method to their own observations; the combined likelihood of a given model would then simply be the product of the likelihoods calculated by each individual experiment.
Near-future gravitational wave detectors are expected to be sensitive to supernovae at distances of up to a few tens of \si{kpc}~\citep{Abbott2018}, making it possible in principle to include the gravitational wave signal to improve the model identification further.

\acknowledgments

We thank MacKenzie Warren, Ken’ichiro Nakazato, Tomonori Totani, Adam Burrows, David Vartanyan and Irene Tamborra for access to the supernova models used in this work and for answering various related questions.

This work was supported by MEXT Grant-in-Aid for Scientific Research on Innovative Areas titled “Exploration of Particle Physics and Cosmology with Neutrinos” under Grants No. 18H05535, No. 18H05536 and No. 18H5537. In addition, participation of individual researchers has been further supported by funds from JSPS, Japan; the European Union’s Horizon 2020 Research and Innovation Programme H2020 grant numbers RISE-GA822070-JENNIFER2 2020 and RISE-GA872549-SK2HK; SSTF-BA1402-06, NRF grants No. 2009-0083526, NRF-2015R1A2A1A05001869, NRF-2016R1D1A1A02936965, NRF-2016R1D1A3B02010606, NRF-2017R1A2B4012757 \added{and NRF-2018R1A6A1A06024970} funded by the Korean government (MSIP); JSPS-RFBR Grant \#20-52-50010/20 and the Ministry of Science and Higher Education under contract \#075-15-2020-778, Russia; Brazilian Funding agencies, CNPq and CAPES; STFC ST/R00031X/2, ST/T002891/1, ST/V002872/1, Consolidated Grants, UKRI MR/S032843/1 and MR/S034102/1, UK.

\vspace{5mm}

\software{BONSAI~\citep{Smy2007},
               sntools~\replaced{\citep{sntools}}{\citep{Migenda2021}},
               WCSim~\citep{WCSim},
               matplotlib~\citep{Hunter2007},
               NumPy~\citep{Walt2011},
               SciPy~\citep{Virtanen2020}
}



\appendix

\section{Derivation of Likelihood Function}\label{apx:likelihood}

In this section, we derive the likelihood function introduced in section~\ref{sec:results-ll}.
It is based on a likelihood function derived by~\citet{Loredo1989} to analyse events from SN1987A, but we extend it to account for multiple interaction channels.

We start by considering bins in time and observed energy, where the bin size $\Delta t \cdot \Delta E$ is sufficiently small that the expected number of events per bin,
\begin{equation}
N_i = \frac{\d^2\,N (E_i, t_i)}{\d E \d t} \Delta E \Delta t,
\end{equation}
is much smaller than 1.
Here, $N(E, t)$ is the observed event rate as a function of time and energy as predicted by a supernova model.

Assuming a Poisson distribution, the probability of observing 0 events in a single interaction channel\footnote{Throughout this appendix, we will use greek letters to refer to interaction channels.} in a bin around time $t_i$ and energy $E_i$ is
\begin{equation}
P_{0, \alpha} = \exp\left( - N_{i, \alpha} \right).
\end{equation}

When considering multiple interaction channels, the probability of observing 0 events is simply the product of the probabilities of observing 0 events in every single interaction channel, i.\,e.
\begin{equation}
P_0 = \prod_\alpha P_{0, \alpha} = \prod_\alpha \exp\left( -N_{i, \alpha} \right).
\end{equation}

In any given interaction channel the probability of observing exactly one event is
\begin{equation}
P_{1, \alpha} = N_{i, \alpha} \exp\left( -N_{i, \alpha} \right)
\end{equation}
and the total probability of observing exactly one event is
\begin{eqnarray}
P_1 &=& \sum_\alpha \left( P_{1, \alpha} \cdot \prod_{\beta \neq \alpha} P_{0, \beta} \right) \\
&=& \sum_\alpha \left[ N_{i, \alpha} \exp\left( -N_{i, \alpha} \right) \cdot \prod_{\beta \neq \alpha} \exp\left( -N_{i, \beta} \right) \right] \\
&=& \sum_\alpha \left[ N_{i, \alpha} \cdot \prod_\beta \exp\left( -N_{i, \beta} \right) \right] \\
&=& P_0 \cdot \sum_\alpha N_{i, \alpha}.
\end{eqnarray}

The bin size was chosen such that the probability of observing more than one event in a bin is negligible.

The likelihood of observing a set of events $(E_i, t_i)$ with $i = 1, …, N_\text{obs}$ is then given by
\begin{eqnarray}
\mathcal{L} &=& \left[ \prod_{i=1}^{N_\text{obs}} P_1 (E_i, t_i)\right] \cdot \prod_{j \neq i} P_0 (E_j, t_j) \\
&=& \left[ \prod_{i=1}^{N_\text{obs}} \left(\sum_\alpha N_{i, \alpha}\right) P_0 (E_i, t_i) \right] \cdot \prod_{j \neq i} P_0 (E_j, t_j) \\
&=& \left[ \prod_{i=1}^{N_\text{obs}} \sum_\alpha N_{i, \alpha} \right] \cdot \prod_j P_0 (E_j, t_j),
\end{eqnarray}
where the products over $j \neq i$ include only bins that do not contain an event, while products over $j$ include all bins.

For simplicity, we consider the log-likelihood $L = \ln \mathcal{L}$.
Using $\ln(a \cdot b) = \ln(a) + \ln(b)$, the log-likelihood function is
\begin{equation}
L = \sum_{i=1}^{N_\text{obs}} \ln \left( \sum_\alpha N_{i, \alpha} \right) + \sum_j \ln \left( P_0 (E_j, t_j) \right),
\end{equation}
where the second term simplifies to
\begin{eqnarray}
\sum_j \ln \left( P_0 (E_j, t_j) \right) &=& \sum_j \ln \left[ \prod_\alpha \exp\left( -N_{j, \alpha} \right) \right] \\
&=& \sum_j \sum_\alpha \ln \left[ \exp\left( -N_{j, \alpha} \right) \right] \\
&=& - \sum_j \sum_\alpha N_{j, \alpha} \\
&=& - N_\text{exp}
\end{eqnarray}
and since we assume in this paper that the distance to the supernova is unknown, we normalize the event rate so as to reproduce the observed number of events.
$N_\text{exp}$ is therefore model-independent and since we only consider likelihood ratios of different models A and B, $\Delta L = L_A - L_B$, this part cancels out.
The final likelihood function we use is therefore given by
\begin{equation}
L = \sum_{i=1}^{N_\text{obs}} \ln \left( \sum_\alpha N_{i, \alpha} \right).
\end{equation}

\section{Pairwise Model Comparisons}\label{apx:model-comparisons}

This appendix contains figures showing pairwise model comparisons\added{, similar to figure~\ref{fig-ana-100normal}}.

\begin{figure}[htbp]
	\centering
	\includegraphics[scale=0.75]{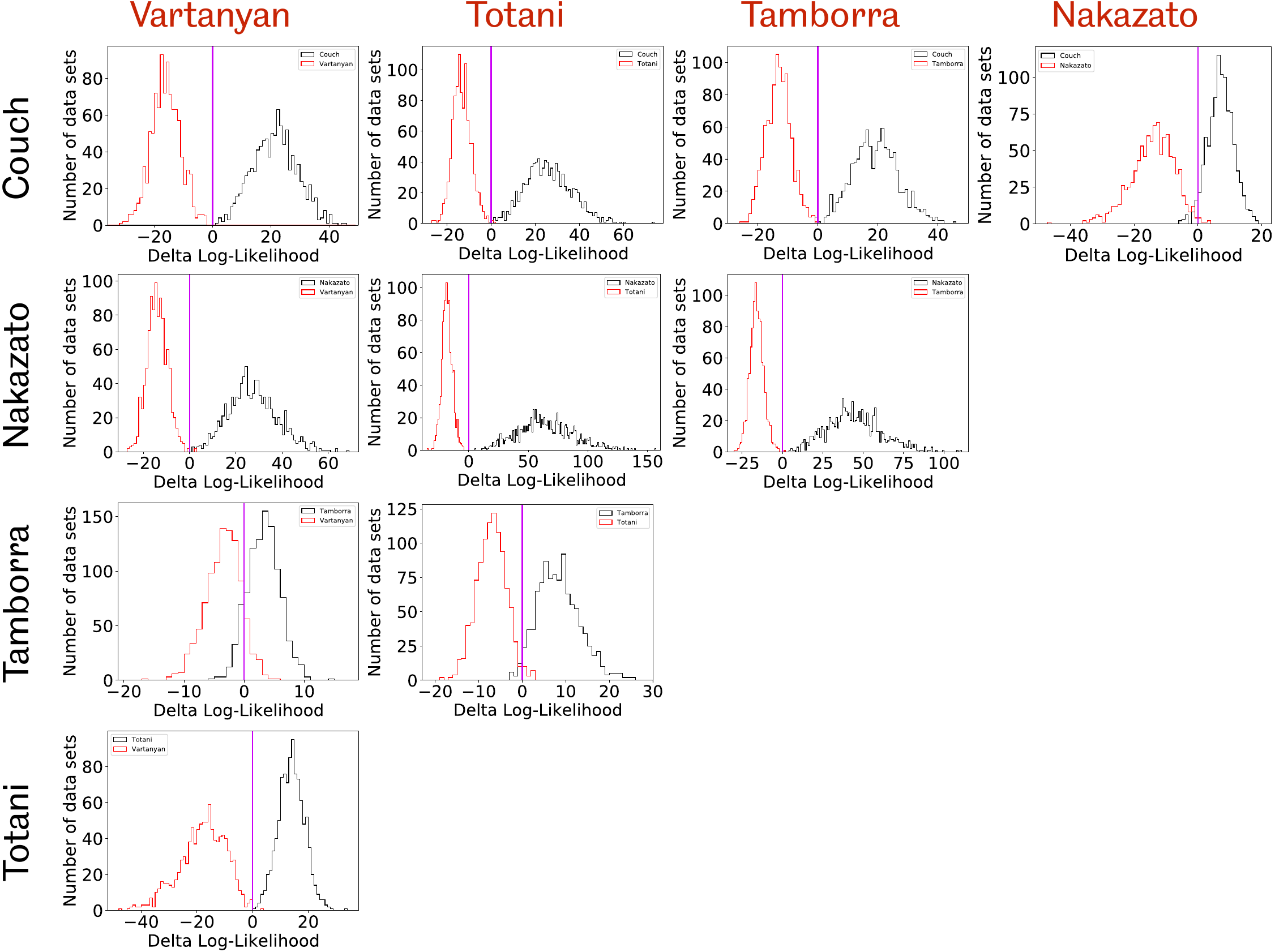}
	\caption{Histograms showing the distribution of $\Delta L = L_\text{black} - L_\text{red}$ for all pairs of supernova models considered here, for 100 events per data set and inverted mass ordering. The purple vertical line in each panel indicates $\Delta L = 0$.}
	\label{fig-ana-100inverted}
\end{figure}

\begin{figure}[htbp]
	\centering
	\includegraphics[scale=0.75]{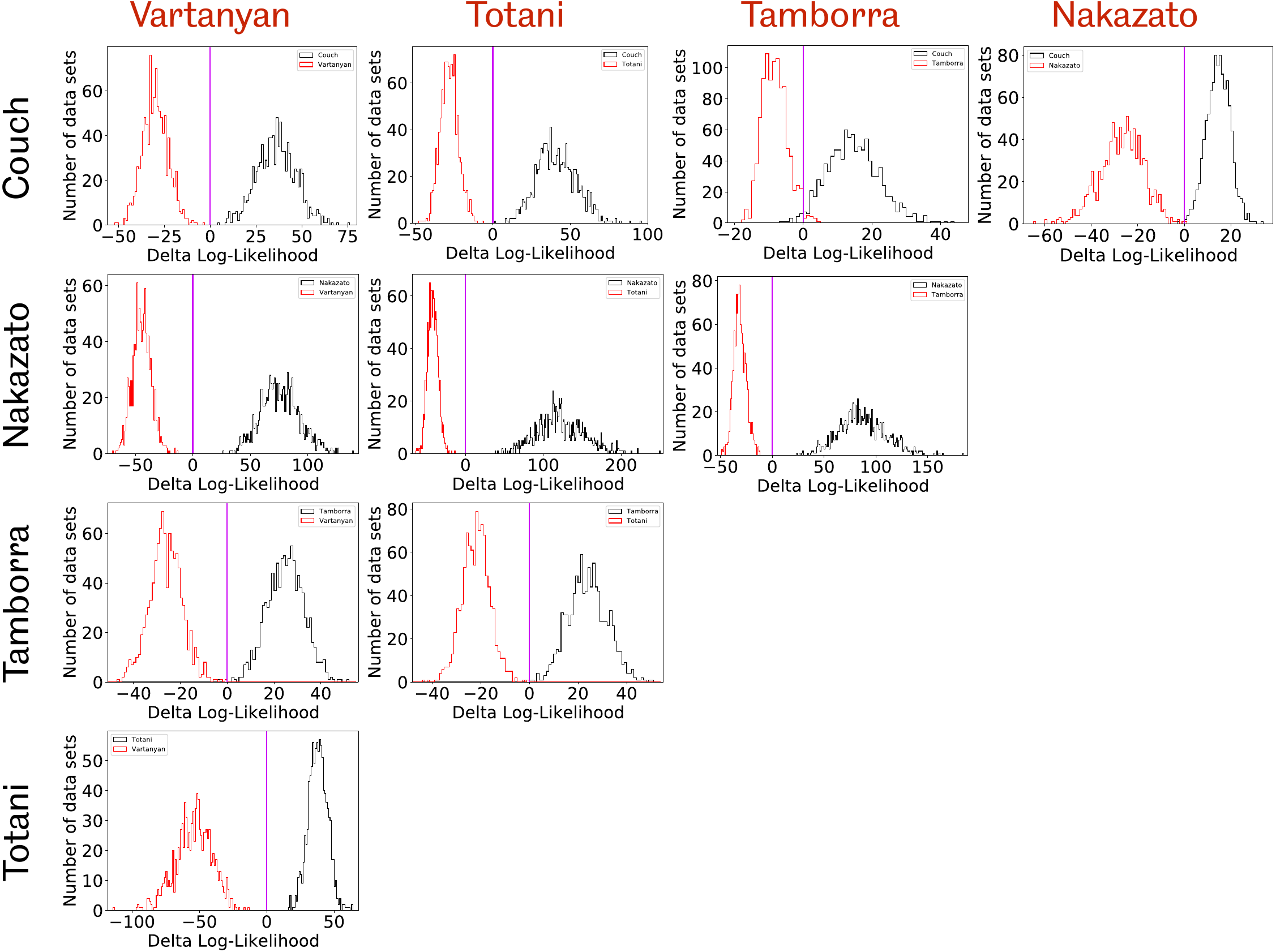}
	\caption{Histograms showing the distribution of $\Delta L = L_\text{black} - L_\text{red}$ for all pairs of supernova models considered here, for 300 events per data set and normal mass ordering. The purple vertical line in each panel indicates $\Delta L = 0$.}
	\label{fig-ana-300normal}
\end{figure}

\begin{figure}[htbp]
	\centering
	\includegraphics[scale=0.75]{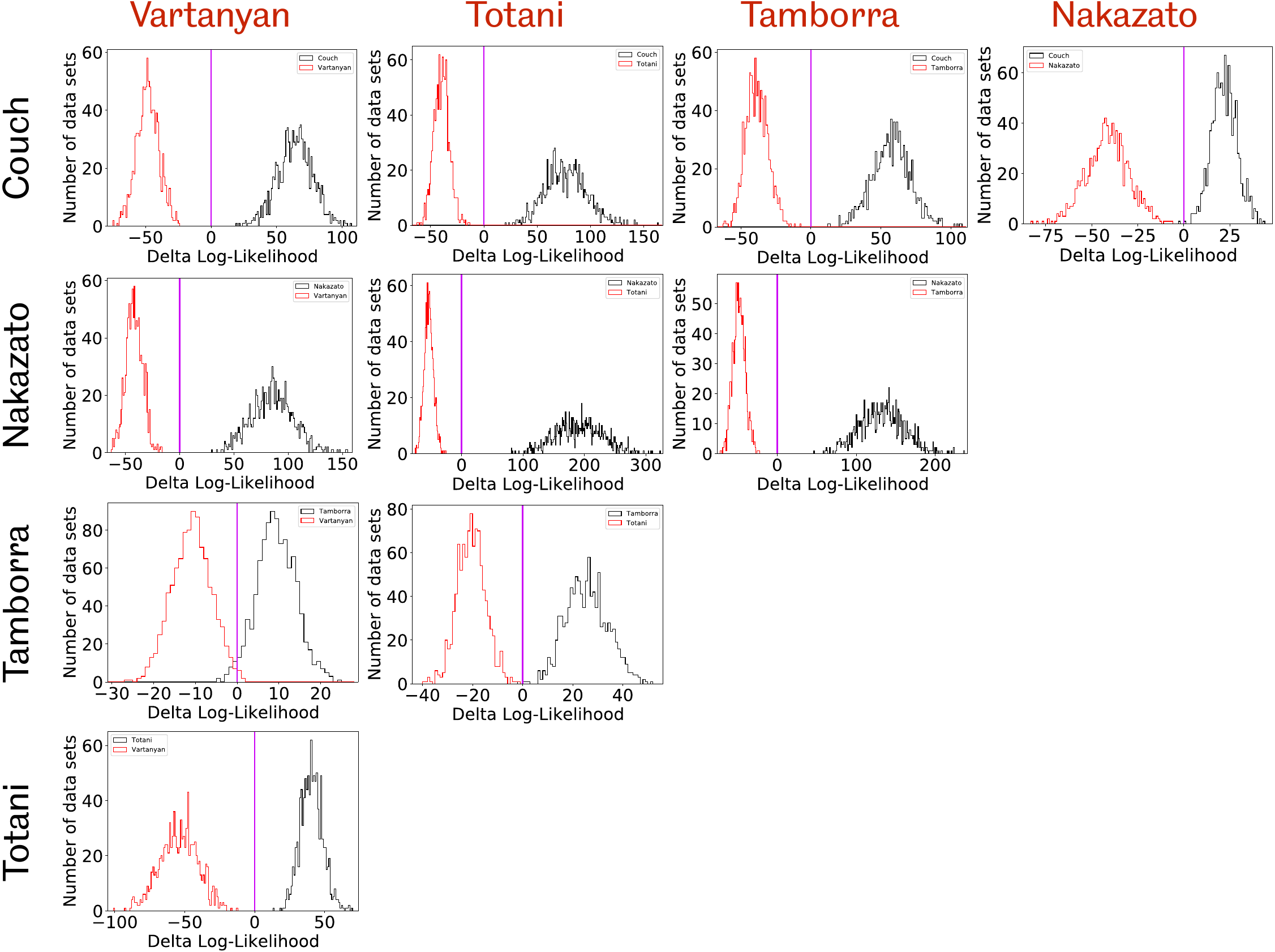}
	\caption{Histograms showing the distribution of $\Delta L = L_\text{black} - L_\text{red}$ for all pairs of supernova models considered here, for 300 events per data set and inverted mass ordering. The purple vertical line in each panel indicates $\Delta L = 0$.}
	\label{fig-ana-300inverted}
\end{figure}


\bibliography{paper}{}

\begin{thebibliography}{}
\expandafter\ifx\csname natexlab\endcsname\relax\def\natexlab#1{#1}\fi
\providecommand{\url}[1]{\href{#1}{#1}}
\providecommand{\dodoi}[1]{doi:~\href{http://doi.org/#1}{\nolinkurl{#1}}}
\providecommand{\doeprint}[1]{\href{http://ascl.net/#1}{\nolinkurl{http://ascl.net/#1}}}
\providecommand{\doarXiv}[1]{\href{https://arxiv.org/abs/#1}{\nolinkurl{https://arxiv.org/abs/#1}}}

\bibitem[{snt(2020)}]{sntools}
 2020, sntools: Event generator for supernova burst neutrinos.
  \url{https://github.com/JostMigenda/sntools}

\bibitem[{WCS(2020)}]{WCSim}
 2020, WCSim: The WCSim GEANT4 application.
  \url{https://github.com/WCSim/WCSim}

\bibitem[{Abbott {et~al.}(2018)}]{Abbott2018}
Abbott, B.~P., {et~al.} 2018, Living Rev. Rel., 21, 3,
  \dodoi{10.1007/s41114-018-0012-9, 10.1007/lrr-2016-1}

\bibitem[{Abe {et~al.}(2011)}]{Abe2011c}
Abe, K., {et~al.} 2011, Phys. Rev., D83, 052010,
  \dodoi{10.1103/PhysRevD.83.052010}

\bibitem[{Abe {et~al.}(2016{\natexlab{a}})}]{Abe2016}
---. 2016{\natexlab{a}}, Astropart. Phys., 81, 39,
  \dodoi{10.1016/j.astropartphys.2016.04.003}

\bibitem[{Abe {et~al.}(2016{\natexlab{b}})}]{Abe2016a}
---. 2016{\natexlab{b}}, Phys. Rev., D94, 052010,
  \dodoi{10.1103/PhysRevD.94.052010}

\bibitem[{Abe {et~al.}(2018)}]{HKDR2018}
---. 2018.
\newblock \doarXiv{1805.04163}

\bibitem[{Abe {et~al.}(2019)}]{Abe:2019fux}
---. 2019.
\newblock \doarXiv{1908.05141}

\bibitem[{Acciarri {et~al.}(2015)}]{Acciarri2015}
Acciarri, R., {et~al.} 2015.
\newblock \doarXiv{1512.06148}

\bibitem[{Adams {et~al.}(2013)Adams, Kochanek, Beacom, Vagins, \&
  Stanek}]{Adams:2013ana}
Adams, S.~M., Kochanek, C.~S., Beacom, J.~F., Vagins, M.~R., \& Stanek, K.~Z.
  2013, Astrophys. J., 778, 164, \dodoi{10.1088/0004-637X/778/2/164}

\bibitem[{Agostinelli {et~al.}(2003)}]{Agostinelli2003}
Agostinelli, S., {et~al.} 2003, Nucl. Instrum. Meth., A506, 250,
  \dodoi{10.1016/S0168-9002(03)01368-8}

\bibitem[{Alekseev {et~al.}(1987)Alekseev, Alekseeva, Volchenko, \&
  Krivosheina}]{Alekseev1987}
Alekseev, E., Alekseeva, L., Volchenko, V., \& Krivosheina, I. 1987, JETP
  Letters, 45, 589

\bibitem[{Bahcall {et~al.}(1995)Bahcall, Kamionkowski, \& Sirlin}]{Bahcall1995}
Bahcall, J.~N., Kamionkowski, M., \& Sirlin, A. 1995, Phys. Rev., D51, 6146,
  \dodoi{10.1103/PhysRevD.51.6146}

\bibitem[{Bethe \& Wilson(1985)}]{Bethe1985}
Bethe, H.~A., \& Wilson, J.~R. 1985, Astrophys. J., 295, 14,
  \dodoi{10.1086/163343}

\bibitem[{Bionta {et~al.}(1987)}]{Bionta1987}
Bionta, R.~M., {et~al.} 1987, Phys. Rev. Lett., 58, 1494,
  \dodoi{10.1103/PhysRevLett.58.1494}

\bibitem[{Brun \& Rademakers(1997)}]{Brun1997}
Brun, R., \& Rademakers, F. 1997, Nucl. Instrum. Meth., A389, 81,
  \dodoi{10.1016/S0168-9002(97)00048-X}

\bibitem[{Buras {et~al.}(2006)Buras, Rampp, Janka, \& Kifonidis}]{Buras2006}
Buras, R., Rampp, M., Janka, H.~T., \& Kifonidis, K. 2006, Astron. Astrophys.,
  447, 1049, \dodoi{10.1051/0004-6361:20053783}

\bibitem[{Burrows {et~al.}(2020)Burrows, Radice, Vartanyan, Nagakura, Skinner,
  \& Dolence}]{Burrows2020}
Burrows, A., Radice, D., Vartanyan, D., {et~al.} 2020, Mon. Not. Roy. Astron.
  Soc., 491, 2715, \dodoi{10.1093/mnras/stz3223}

\bibitem[{Burrows {et~al.}(2006)Burrows, Reddy, \& Thompson}]{Burrows2006}
Burrows, A., Reddy, S., \& Thompson, T.~A. 2006, Nucl. Phys., A777, 356,
  \dodoi{10.1016/j.nuclphysa.2004.06.012}

\bibitem[{Chakraborty {et~al.}(2016)Chakraborty, Hansen, Izaguirre, \&
  Raffelt}]{Chakraborty2016}
Chakraborty, S., Hansen, R., Izaguirre, I., \& Raffelt, G. 2016, Nucl. Phys.,
  B908, 366, \dodoi{10.1016/j.nuclphysb.2016.02.012}

\bibitem[{Colella \& Woodward(1984)}]{Colella1984}
Colella, P., \& Woodward, P.~R. 1984, J. Comput. Phys., 54, 174,
  \dodoi{10.1016/0021-9991(84)90143-8}

\bibitem[{Couch {et~al.}(2019)Couch, Warren, \& O'Connor}]{Couch2019}
Couch, S.~M., Warren, M.~L., \& O'Connor, E.~P. 2019.
\newblock \doarXiv{1902.01340}

\bibitem[{Dighe \& Smirnov(2000)}]{Dighe2000}
Dighe, A.~S., \& Smirnov, A.~{\relax Yu}. 2000, Phys. Rev., D62, 033007,
  \dodoi{10.1103/PhysRevD.62.033007}

\bibitem[{Duan {et~al.}(2006{\natexlab{a}})Duan, Fuller, Carlson, \&
  Qian}]{Duan2006a}
Duan, H., Fuller, G.~M., Carlson, J., \& Qian, Y.-Z. 2006{\natexlab{a}}, Phys.
  Rev. Lett., 97, 241101, \dodoi{10.1103/PhysRevLett.97.241101}

\bibitem[{Duan {et~al.}(2006{\natexlab{b}})Duan, Fuller, Carlson, \&
  Qian}]{Duan2006}
---. 2006{\natexlab{b}}, Phys. Rev., D74, 105014,
  \dodoi{10.1103/PhysRevD.74.105014}

\bibitem[{Dubey {et~al.}(2009)Dubey, Antypas, Ganapathy, Reid, Riley, Sheeler,
  Siegel, \& Weide}]{Dubey2009}
Dubey, A., Antypas, K., Ganapathy, M.~K., {et~al.} 2009, Parallel Computing,
  35, 512 , \dodoi{https://doi.org/10.1016/j.parco.2009.08.001}

\bibitem[{Fogli {et~al.}(2005)Fogli, Lisi, Mirizzi, \& Montanino}]{Fogli2005}
Fogli, G.~L., Lisi, E., Mirizzi, A., \& Montanino, D. 2005, JCAP, 0504, 002,
  \dodoi{10.1088/1475-7516/2005/04/002}

\bibitem[{Fryxell {et~al.}(1991)Fryxell, Müller, \& Arnett}]{Fryxell1991}
Fryxell, B., Müller, E., \& Arnett, W.~D. 1991, Astrophys. J., 367, 619,
  \dodoi{10.1086/169657}

\bibitem[{Fryxell {et~al.}(2000)Fryxell, Olson, Ricker, Timmes, Zingale, Lamb,
  MacNeice, Rosner, Truran, \& Tufo}]{Fryxell2000}
Fryxell, B., Olson, K., Ricker, P., {et~al.} 2000, Astrophys. J. Suppl., 131,
  273, \dodoi{10.1086/317361}

\bibitem[{Fukuda {et~al.}(2003)}]{Fukuda2003}
Fukuda, S., {et~al.} 2003, Nucl. Instrum. Meth., A501, 418

\bibitem[{Hanke {et~al.}(2013)Hanke, Mueller, Wongwathanarat, Marek, \&
  Janka}]{Hanke2013}
Hanke, F., Mueller, B., Wongwathanarat, A., Marek, A., \& Janka, H.-T. 2013,
  Astrophys. J., 770, 66.
\newblock \url{http://arxiv.org/abs/1303.6269}

\bibitem[{Hilditch {et~al.}(2005)Hilditch, Howarth, \& Harries}]{Hilditch2005}
Hilditch, R.~W., Howarth, I.~D., \& Harries, T.~J. 2005, Mon. Not. Roy. Astron.
  Soc., 357, 304, \dodoi{10.1111/j.1365-2966.2005.08653.x}

\bibitem[{Hirata {et~al.}(1987)}]{Hirata1987}
Hirata, K.~S., {et~al.} 1987, Phys. Rev. Lett., 58, 1490,
  \dodoi{10.1103/PhysRevLett.58.1490}

\bibitem[{Hirata {et~al.}(1988)}]{Hirata1988a}
---. 1988, Phys. Rev., D38, 448, \dodoi{10.1103/PhysRevD.38.448}

\bibitem[{Horiuchi \& Kneller(2018)}]{Horiuchi2018}
Horiuchi, S., \& Kneller, J.~P. 2018, J. Phys., G45, 043002,
  \dodoi{10.1088/1361-6471/aaa90a}

\bibitem[{Horiuchi {et~al.}(2017)Horiuchi, Nakamura, Takiwaki, \&
  Kotake}]{Horiuchi2017}
Horiuchi, S., Nakamura, K., Takiwaki, T., \& Kotake, K. 2017, J. Phys., G44,
  114001, \dodoi{10.1088/1361-6471/aa8f1f}

\bibitem[{Hunter(2007)}]{Hunter2007}
Hunter, J.~D. 2007, Comput. Sci. Eng., 9, 90, \dodoi{10.1109/MCSE.2007.55}

\bibitem[{Ianni {et~al.}(2009)Ianni, Pagliaroli, Strumia, Torres, Villante, \&
  Vissani}]{Ianni2009}
Ianni, A., Pagliaroli, G., Strumia, A., {et~al.} 2009, Phys. Rev., D80, 043007,
  \dodoi{10.1103/PhysRevD.80.043007}

\bibitem[{Kachelrieß {et~al.}(2005)Kachelrieß, Tomàs, Buras, Janka, Marek,
  \& Rampp}]{Kachelries2005}
Kachelrieß, M., Tomàs, R., Buras, R., {et~al.} 2005, Phys. Rev., D71

\bibitem[{Kass \& Raftery(1995)}]{Kass1995}
Kass, R.~E., \& Raftery, A.~E. 1995, J. Am. Statist. Assoc., 90, 773,
  \dodoi{10.1080/01621459.1995.10476572}

\bibitem[{Kurylov {et~al.}(2003)Kurylov, Ramsey-Musolf, \& Vogel}]{Kurylov2003}
Kurylov, A., Ramsey-Musolf, M.~J., \& Vogel, P. 2003, Phys. Rev., C67, 035502,
  \dodoi{10.1103/PhysRevC.67.035502}

\bibitem[{Langanke {et~al.}(1996)Langanke, Vogel, \& Kolbe}]{Langanke:1995he}
Langanke, K., Vogel, P., \& Kolbe, E. 1996, Phys. Rev. Lett., 76, 2629,
  \dodoi{10.1103/PhysRevLett.76.2629}

\bibitem[{Lattimer \& Swesty(1991)}]{Lattimer1991}
Lattimer, J.~M., \& Swesty, F.~D. 1991, Nucl. Phys., A535, 331,
  \dodoi{10.1016/0375-9474(91)90452-C}

\bibitem[{Li {et~al.}(2021)Li, Roberts, \& Beacom}]{Li:2020ujl}
Li, S.~W., Roberts, L.~F., \& Beacom, J.~F. 2021, Phys. Rev. D, 103, 023016,
  \dodoi{10.1103/PhysRevD.103.023016}

\bibitem[{Loredo \& Lamb(1989)}]{Loredo1989}
Loredo, T.~J., \& Lamb, D.~Q. 1989, Annals N. Y. Acad. Sci., 571, 601,
  \dodoi{10.1111/j.1749-6632.1989.tb50547.x}

\bibitem[{Loredo \& Lamb(2002)}]{Loredo2002}
---. 2002, Phys. Rev., D65, 063002, \dodoi{10.1103/PhysRevD.65.063002}

\bibitem[{Lund {et~al.}(2010)Lund, Marek, Lunardini, Janka, \&
  Raffelt}]{Lund2010}
Lund, T., Marek, A., Lunardini, C., Janka, H.-T., \& Raffelt, G. 2010, Phys.
  Rev., D82, 063007, \dodoi{10.1103/PhysRevD.82.063007}

\bibitem[{Marek {et~al.}(2006)Marek, Dimmelmeier, Janka, Müller, \&
  Buras}]{Marek2006}
Marek, A., Dimmelmeier, H., Janka, H.~T., Müller, E., \& Buras, R. 2006,
  Astron. Astrophys., 445, 273, \dodoi{10.1051/0004-6361:20052840}

\bibitem[{Mayle {et~al.}(1987)Mayle, Wilson, \& Schramm}]{Mayle1987}
Mayle, R., Wilson, J.~R., \& Schramm, D.~N. 1987, Astrophys. J., 318, 288,
  \dodoi{10.1086/165367}

\bibitem[{Migenda {et~al.}(2021)Migenda, Cartwright, Kneale, Malek,
  Schnellbach, \& Stone}]{Migenda2021}
Migenda, J., Cartwright, S., Kneale, L., {et~al.} 2021, J. of Open Source
  Software, \dodoi{10.21105/joss.2877}

\bibitem[{Mirizzi {et~al.}(2016)Mirizzi, Tamborra, Janka, Saviano, Scholberg,
  Bollig, Hudepohl, \& Chakraborty}]{Mirizzi:2015eza}
Mirizzi, A., Tamborra, I., Janka, H.-T., {et~al.} 2016, Riv. Nuovo Cim., 39, 1,
  \dodoi{10.1393/ncr/i2016-10120-8}

\bibitem[{Murchikova {et~al.}(2017)Murchikova, Abdikamalov, \&
  Urbatsch}]{Murchikova2017}
Murchikova, L.~M., Abdikamalov, E., \& Urbatsch, T. 2017, Mon. Not. Roy.
  Astron. Soc., 469, 1725, \dodoi{10.1093/mnras/stx986}

\bibitem[{Müller {et~al.}(2012)Müller, Janka, \& Marek}]{Muller2012}
Müller, B., Janka, H.-T., \& Marek, A. 2012, Astrophys. J., 756, 84

\bibitem[{Nakazato {et~al.}(2013)Nakazato, Sumiyoshi, Suzuki, Totani, Umeda, \&
  Yamada}]{Nakazato2013}
Nakazato, K., Sumiyoshi, K., Suzuki, H., {et~al.} 2013, Astrophys. J. Suppl.,
  205, 2, \dodoi{10.1088/0067-0049/205/1/2}

\bibitem[{Nakazato \& Suzuki(2020)}]{Nakazato:2020ogl}
Nakazato, K., \& Suzuki, H. 2020, Astrophys. J., 891, 156,
  \dodoi{10.3847/1538-4357/ab7456}

\bibitem[{Nakazato {et~al.}(2018)Nakazato, Suzuki, \& Sakuda}]{Nakazato2018}
Nakazato, K., Suzuki, T., \& Sakuda, M. 2018, PTEP, 2018, 123E02,
  \dodoi{10.1093/ptep/pty134}

\bibitem[{Nikrant {et~al.}(2018)Nikrant, Laha, \& Horiuchi}]{Nikrant2018}
Nikrant, A., Laha, R., \& Horiuchi, S. 2018, Phys. Rev., D97, 023019,
  \dodoi{10.1103/PhysRevD.97.023019}

\bibitem[{O'Connor(2015)}]{OConnor2015}
O'Connor, E. 2015, Astrophys. J. Suppl., 219, 24,
  \dodoi{10.1088/0067-0049/219/2/24}

\bibitem[{O'Connor \& Couch(2018{\natexlab{a}})}]{OConnor2018}
O'Connor, E.~P., \& Couch, S.~M. 2018{\natexlab{a}}, Astrophys. J., 865, 81,
  \dodoi{10.3847/1538-4357/aadcf7}

\bibitem[{O'Connor \& Couch(2018{\natexlab{b}})}]{OConnor2018b}
---. 2018{\natexlab{b}}, Astrophys. J., 854, 63,
  \dodoi{10.3847/1538-4357/aaa893}

\bibitem[{Pietrzyński {et~al.}(2013)}]{Pietrzynski2013}
Pietrzyński, G., {et~al.} 2013, Nature, 495, 76, \dodoi{10.1038/nature11878}

\bibitem[{Radice {et~al.}(2017)Radice, Burrows, Vartanyan, Skinner, \&
  Dolence}]{Radice2017}
Radice, D., Burrows, A., Vartanyan, D., Skinner, M.~A., \& Dolence, J.~C. 2017,
  Astrophys. J., 850, 43, \dodoi{10.3847/1538-4357/aa92c5}

\bibitem[{Rampp \& Janka(2002)}]{Rampp2002}
Rampp, M., \& Janka, H.-T. 2002, Astron. Astrophys., 396, 361

\bibitem[{Scholberg(2012)}]{Scholberg:2012id}
Scholberg, K. 2012, Ann. Rev. Nucl. Part. Sci., 62, 81,
  \dodoi{10.1146/annurev-nucl-102711-095006}

\bibitem[{Seadrow {et~al.}(2018)Seadrow, Burrows, Vartanyan, Radice, \&
  Skinner}]{Seadrow2018}
Seadrow, S., Burrows, A., Vartanyan, D., Radice, D., \& Skinner, M.~A. 2018,
  Mon. Not. Roy. Astron. Soc., 480, 4710, \dodoi{10.1093/mnras/sty2164}

\bibitem[{Shen {et~al.}(1998)Shen, Toki, Oyamatsu, \& Sumiyoshi}]{Shen1998}
Shen, H., Toki, H., Oyamatsu, K., \& Sumiyoshi, K. 1998, Prog. Theor. Phys.,
  100, 1013, \dodoi{10.1143/PTP.100.1013}

\bibitem[{Shibata {et~al.}(2011)Shibata, Kiuchi, Sekiguchi, \&
  Suwa}]{Shibata2011}
Shibata, M., Kiuchi, K., Sekiguchi, Y.-i., \& Suwa, Y. 2011, Prog. Theor.
  Phys., 125, 1255, \dodoi{10.1143/PTP.125.1255}

\bibitem[{Skinner {et~al.}(2019)Skinner, Dolence, Burrows, Radice, \&
  Vartanyan}]{Skinner2019}
Skinner, M.~A., Dolence, J.~C., Burrows, A., Radice, D., \& Vartanyan, D. 2019,
  Astrophys. J. Suppl., 241, 7, \dodoi{10.3847/1538-4365/ab007f}

\bibitem[{Smy(2007)}]{Smy2007}
Smy, M. 2007, in {Proceedings, 30th International Cosmic Ray Conference (ICRC
  2007): Merida, Yucatan, Mexico, July 3-11, 2007}, Vol.~5, 1279--1282.
\newblock
  \url{http://indico.nucleares.unam.mx/contributionDisplay.py?contribId=213&confId=4}

\bibitem[{Steiner {et~al.}(2013)Steiner, Hempel, \& Fischer}]{Steiner2013}
Steiner, A.~W., Hempel, M., \& Fischer, T. 2013, Astrophys. J., 774, 17,
  \dodoi{10.1088/0004-637X/774/1/17}

\bibitem[{Strumia \& Vissani(2003)}]{Strumia2003}
Strumia, A., \& Vissani, F. 2003, Phys. Lett., B564, 42

\bibitem[{Sukhbold {et~al.}(2016)Sukhbold, Ertl, Woosley, Brown, \&
  Janka}]{Sukhbold2016}
Sukhbold, T., Ertl, T., Woosley, S.~E., Brown, J.~M., \& Janka, H.~T. 2016,
  Astrophys. J., 821, 38, \dodoi{10.3847/0004-637X/821/1/38}

\bibitem[{Sukhbold \& Woosley(2014)}]{Sukhbold2014}
Sukhbold, T., \& Woosley, S. 2014, Astrophys. J., 783, 10,
  \dodoi{10.1088/0004-637X/783/1/10}

\bibitem[{Sumiyoshi {et~al.}(2005)Sumiyoshi, Yamada, Suzuki, Shen, Chiba, \&
  Toki}]{Sumiyoshi2005}
Sumiyoshi, K., Yamada, S., Suzuki, H., {et~al.} 2005, Astrophys. J., 629, 922,
  \dodoi{10.1086/431788}

\bibitem[{Suwa {et~al.}(2019)Suwa, Sumiyoshi, Nakazato, Takahira, Koshio, Mori,
  \& Wendell}]{Suwa2019}
Suwa, Y., Sumiyoshi, K., Nakazato, K., {et~al.} 2019, Astrophys. J., 881, 139,
  \dodoi{10.3847/1538-4357/ab2e05}

\bibitem[{Suzuki {et~al.}(2018)Suzuki, Chiba, Yoshida, Takahashi, \&
  Umeda}]{Suzuki2018}
Suzuki, T., Chiba, S., Yoshida, T., Takahashi, K., \& Umeda, H. 2018, Phys.
  Rev., C98, 034613, \dodoi{10.1103/PhysRevC.98.034613}

\bibitem[{Tamborra {et~al.}(2013)Tamborra, Hanke, Müller, Janka, \&
  Raffelt}]{Tamborra2013}
Tamborra, I., Hanke, F., Müller, B., Janka, H.-T., \& Raffelt, G. 2013, Phys.
  Rev. Lett., 111, 121104, \dodoi{10.1103/PhysRevLett.111.121104}

\bibitem[{Tamborra {et~al.}(2014)Tamborra, Raffelt, Hanke, Janka, \&
  Mueller}]{Tamborra2014}
Tamborra, I., Raffelt, G., Hanke, F., Janka, H.-T., \& Mueller, B. 2014, Phys.
  Rev., D90, 045032, \dodoi{10.1103/PhysRevD.90.045032}

\bibitem[{Tanabashi {et~al.}(2018)}]{PDG2018}
Tanabashi, M., {et~al.} 2018, Phys. Rev., D98,
  \dodoi{10.1103/PhysRevD.98.030001}

\bibitem[{{Thorne}(1981)}]{Thorne1981}
{Thorne}, K.~S. 1981, Mon. Not. Roy. Astron. Soc., 194, 439,
  \dodoi{10.1093/mnras/194.2.439}

\bibitem[{Totani {et~al.}(1998)Totani, Sato, Dalhed, \& Wilson}]{Totani1998}
Totani, T., Sato, K., Dalhed, H.~E., \& Wilson, J.~R. 1998, Astrophys. J.,
  496, 216, \dodoi{10.1086/305364}

\bibitem[{van~der Walt {et~al.}(2011)van~der Walt, Colbert, \&
  Varoquaux}]{Walt2011}
van~der Walt, S., Colbert, S.~C., \& Varoquaux, G. 2011, Comput. Sci. Eng., 13,
  22, \dodoi{10.1109/MCSE.2011.37}

\bibitem[{{Virtanen} {et~al.}(2020){Virtanen}, {Gommers}, {Oliphant},
  {Haberland}, {Reddy}, {Cournapeau}, {Burovski}, {Peterson}, {Weckesser},
  {Bright}, {van der Walt}, {Brett}, {Wilson}, {Jarrod Millman}, {Mayorov},
  {Nelson}, {Jones}, {Kern}, {Larson}, {Carey}, {Polat}, {Feng}, {Moore}, {Vand
  erPlas}, {Laxalde}, {Perktold}, {Cimrman}, {Henriksen}, {Quintero}, {Harris},
  {Archibald}, {Ribeiro}, {Pedregosa}, {van Mulbregt}, \&
  {Contributors}}]{Virtanen2020}
{Virtanen}, P., {Gommers}, R., {Oliphant}, T.~E., {et~al.} 2020, Nature
  Methods, 17, 261, \dodoi{https://doi.org/10.1038/s41592-019-0686-2}

\bibitem[{Vissani(2015)}]{Vissani:2014doa}
Vissani, F. 2015, J. Phys. G, 42, 013001, \dodoi{10.1088/0954-3899/42/1/013001}

\bibitem[{Warren(2019)}]{Warren2019}
Warren, M.~L. 2019, {Test upload of neutrino emission from \SI{20}{\Msol}
  progenitor}, 0.2,  Zenodo, \dodoi{10.5072/zenodo.257869}

\bibitem[{Wilson(1985)}]{Wilson1982}
Wilson, J.~R. 1985, in Numerical Astrophysics. Proceedings of the Symposium in
  honour of James R. Wilson, held at the University of Illinois Urbana
  Champaign, October, 1982, ed. J.~M. Centrella, J.~M. LeBlanc, \& R.~L. Bowers
  (Boston: Jones and Bartlett Publ.), 422

\bibitem[{Wilson {et~al.}(1986)Wilson, Mayle, Woosley, \& Weaver}]{Wilson1986}
Wilson, J.~R., Mayle, R., Woosley, S.~E., \& Weaver, T. 1986, Annals N. Y.
  Acad. Sci., 470, 267, \dodoi{10.1111/j.1749-6632.1986.tb47980.x}

\bibitem[{Woosley {et~al.}(2002)Woosley, Heger, \& Weaver}]{Woosley2002}
Woosley, S.~E., Heger, A., \& Weaver, T.~A. 2002, Rev. Mod. Phys., 74, 1015

\end{thebibliography}
\bibliographystyle{aasjournal}



\end{document}